\PassOptionsToPackage{pdfpagelabels=false}{hyperref}
\documentclass[fleqn,usenatbib,useAMS]{mnras}

\usepackage{graphicx}
\usepackage{amssymb}
\usepackage{amsmath}
\usepackage[dvipsnames]{xcolor}
\usepackage[figure,figure*,table,table*]{hypcap}
\usepackage{subfig}
\usepackage{multirow}
\usepackage{ulem}

\newcommand{\hgpc}{h^{-1} \, \mathrm{Gpc}}
\newcommand{\hmpc}{h^{-1} \, \mathrm{Mpc}}
\newcommand{\hkpc}{h^{-1} \, \mathrm{kpc}}
\newcommand{\rp}{r_p}
\renewcommand{\wp}{w_p}

\title[Full-scale and full-shape analysis of RSD and GGL]{Constraints on $S_8$ from a full-scale and full-shape analysis of redshift-space clustering and galaxy--galaxy lensing in BOSS}
\author[J. U. Lange et al.]{Johannes U. Lange$^{1, 2, 3, 4}$\thanks{email: julange.astro@pm.me}, Andrew P. Hearin$^5$, Alexie Leauthaud$^2$, Frank C. van den Bosch$^6$,\newauthor Enia Xhakaj$^2$, Hong Guo$^7$, Risa H. Wechsler$^1$ and Joseph DeRose$^8$\\
$^1$Kavli Institute for Particle Astrophysics and Cosmology and Department of Physics, Stanford University, CA 94305, USA\\
$^2$Department of Astronomy and Astrophysics, University of California, Santa Cruz, CA 95064, USA\\
$^3$Department of Physics, University of Michigan, Ann Arbor, MI 48109, USA\\
$^4$Leinweber Center for Theoretical Physics, University of Michigan, Ann Arbor, MI 48109, USA\\
$^5$Argonne National Laboratory, Argonne, IL 60439, USA\\
$^6$Department of Astronomy, Yale University, New Haven, CT 06511, USA\\
$^7$Key Laboratory for Research in Galaxies and Cosmology, Shanghai Astronomical Observatory, Shanghai 200030, China\\
$^8$Berkeley Center for Cosmological Physics, University of California, Berkeley, CA 94720, USA}
\pubyear{\the\year{}}

\begin{document}

\date{Accepted xxx. Received xxx}

\label{firstpage}
\pagerange{\pageref{firstpage}--\pageref{lastpage}}

\maketitle

\begin{abstract}
    We present a novel simulation-based cosmological analysis of galaxy--galaxy lensing and galaxy redshift-space clustering. Compared to analysis methods based on perturbation theory, our simulation-based approach allows us to probe a much wider range of scales, $0.4 \, \hmpc$ to $63 \, \hmpc$, including highly non-linear scales, and marginalises over astrophysical effects such as assembly bias. We apply this framework to data from the \textit{Baryon Oscillation Spectroscopic Survey} LOWZ sample cross-correlated with state-of-the-art gravitational lensing catalogues from the \textit{Kilo Degree Survey} and the \textit{Dark Energy Survey}. We show that  gravitational lensing and redshift-space clustering when analysed over a large range of scales place tight constraints on the growth-of-structure parameter $S_8 = \sigma_8 \sqrt{\Omega_{\rm m} / 0.3}$. Overall, we infer $S_8 = 0.792 \pm 0.022$ when analysing the combination of galaxy--galaxy lensing and projected galaxy clustering and $S_8 = 0.771 \pm 0.027$ for galaxy redshift-space clustering. These findings highlight the potential constraining power of full-scale studies over studies analysing only large scales, and also showcase the benefits of analysing multiple large-scale structure surveys jointly. Our inferred values for $S_8$ fall below the value inferred from the CMB, $S_8 = 0.834 \pm 0.016$. While this difference is not statistically significant by itself, our results mirror other findings in the literature whereby low-redshift large scale structure probes infer lower values for $S_8$ than the CMB, the so-called $S_8$-tension.
\end{abstract}


\section{Introduction}

The large-scale structure (LSS) distribution in the low-redshift Universe has emerged as one of the primary probes of cosmology. LSS surveys such as the \textit{Baryon Oscillation Spectroscopic Survey} \citep[BOSS; ][]{Reid2016_MNRAS_455_1553, Ahumada2020_ApJS_249_3}, the \textit{Kilo Degree Survey} \citep[KiDS; ][]{Giblin2021_AA_645_105} and the \textit{Dark Energy Survey} \citep[DES; ][]{Gatti2021_MNRAS_504_4312} have provided some of the most stringent constraints on the parameters of the cosmological standard model. In the coming decade, LSS surveys such as the \textit{Dark Energy Spectroscopic Instrument} \citep[DESI; ][]{Abareshi2022_AJ_164_207} survey, the \textit{Legacy Survey of Space and Time} \citep[LSST; ][]{TheLSSTDarkEnergyScienceCollaboration2018_arXiv_1809_1669} on the Vera C. Rubin Observatory, and the Nancy Grace Roman Space Telescope \citep{Spergel2015_arXiv_1503_3757} will build upon this success and provide even more powerful constraints on the cosmological model. LSS surveys are particularly sensitive to $\Omega_{\rm m}$, the fraction of the energy density in matter, and $\sigma_8$, the amplitude of matter fluctuations on scales of $8 \, \hmpc$. There are two promising avenues to constrain $\Omega_{\rm m}$ and $\sigma_8$ from LSS observations: the deflection of light, so-called gravitational lensing, by the LSS mass distribution and the clustering of matter in redshift-space which is sensitive to peculiar velocities via redshift-space distortions (RSDs).

Recently, several LSS probes of the low-redshift Universe have reported tensions with respect to cosmological parameters preferred by the analysis of the high-redshift cosmic microwave background (CMB) under the canonical $\Lambda$ Cold Dark Matter ($\Lambda$CDM) model. Most often this tension is expressed in constraints on the cosmological parameter $S_8 = \sigma_8 \sqrt{\Omega_{\rm m} / 0.3}$ and, hence, is called the ``$S_8$-tension''. Similar to the well-known Hubble tension \citep{Knox2020_PhRvD_101_3533}, the $S_8$-tension could potentially point to new physics beyond the standard $\Lambda$CDM cosmological model. Under $\Lambda$CDM, observations of the CMB by \citet[Planck2020, ][]{PlanckCollaboration2020_AA_641_6} infer $S_8 = 0.834 \pm 0.016$ (TT,TE,EE+lowE), a value that is $5-10\%$ higher than the value preferred by LSS data. Currently, the tension is $2 - 4 \sigma$ significant with respect to several LSS studies utilising gravitational lensing, while the significance is lower for redshift-space clustering studies \citep[see][for a review]{Abdalla2022_JHEAp_34_49}. While no study by itself can claim a $> 5 \sigma$ tension between the low-redshift LSS and the high-redshift CMB, the similarity of the findings of multiple, independent LSS surveys using different techniques suggest that the $S_8$-tension might be a genuine cosmological tension.

Most often, cosmology studies of the LSS distribution focus on large scales where the statistical properties of the matter distribution can be calculated analytically. Furthermore, on large scales, the relationship between the observed galaxy distribution and the underlying dark matter density field can be characterised by a small number of bias factors. However, methods applicable to large, linear scales typically break down on highly non-linear scales. In one approach to extending LSS predictions to small scales, an analytical halo model is used to make predictions for basic summary statistics of the density field such as the matter power spectrum, and the abundance and clustering of dark matter halos \citep{Seljak2000_MNRAS_318_203}. When augmented with additional ingredients for the halo occupation statistics of galaxies, the halo model becomes a prediction pipeline for the galaxy distribution on non-linear scales \citep{Cooray2002_PhR_372_1, vandenBosch2013_MNRAS_430_725, Krause2017_MNRAS_470_2100}. One of the biggest challenges faced by this approach is meeting the stringent demands for percent-level accuracy with an analytic model \citep[e.g.,][]{Tinker2008_ApJ_688_709, Hayashi2008_MNRAS_388_2, Fedeli2014_JCAP_08_028, Miyatake2022_PhRvD_106_3519, Mahony2022_MNRAS_515_2612}; however, steady improvements have been made over the last decade \citep{Mead2015_MNRAS_454_1958, Garcia2021_MNRAS_505_1195, Mead2021_MNRAS_502_1401}, and by now numerous analyses have used such analytical methods to derive constraints on cosmology from LSS measurements in the non-linear regime \citep{Cacciato2013_MNRAS_430_767, Reddick2014_ApJ_783_118, Troster2022_AA_660_27}. We will generically refer to this type of analysis as a ``full-scale'' study, in contrast to analyses that restrict attention to large scales only \citep[see also][for the term ``full-shape'']{Brieden2021_JCAP_12_054}.

In recent years, there has been tremendous progress in computational power and statistical methods \citep{Parejko2013_MNRAS_429_98, Kwan2015_ApJ_810_35, DeRose2019_ApJ_875_69, Zhai2019_ApJ_874_95, Lange2019_MNRAS_490_1870, Nishimichi2019_ApJ_884_29, Wibking2019_MNRAS_484_989, Wibking2020_MNRAS_492_2872, Lange2022_MNRAS_509_1779, Miyatake2022_PhRvD_106_3520} such that one can now directly compare predictions from simulations against LSS observations for multiple cosmological models and perform a rigorous Bayesian quantification of cosmological parameters. These simulation-based methods differ from conventional analytical halo models in that cosmological simulations are {\it directly} populated with synthetic galaxies, and summary statistics are predicted using statistical estimators of the resulting synthetic data. These simulation-based approaches simplify the task of incorporating systematic effects such as galaxy assembly bias \citep[e.g.,][]{Hearin2016_MNRAS_460_2552} and velocity bias \citep[e.g.,][]{Guo2015_MNRAS_446_578}, and at the same time enable a full-scale analysis to obtain stringent cosmological constraints \citep{Wibking2019_MNRAS_484_989, Zhai2019_ApJ_874_95, Lange2022_MNRAS_509_1779, Salcedo2022_MNRAS_510_5376}. Thus, simulation-based full-scale studies have potential to exhaust the information content of LSS two-point correlation functions using analyses with a degree of complexity that is difficult to achieve with an analytical model.

Simulation-based full-scale studies have matured in recent years to the point that several have been applied to actual LSS data, demonstrating that the long-forecasted constraining power of full-scale studies can be realised. For example, \cite{Wibking2020_MNRAS_492_2872} use a full-scale approach to infer $S_8 \approx 0.712 \pm 0.031$ from a combination of galaxy clustering and galaxy--galaxy lensing. In a similar study using updated lensing data from the \textit{Hyper Suprime-Cam} (HSC) survey, \cite{Miyatake2022_PhRvD_106_3520} infer $S_8 = 0.795_{-0.042}^{+0.049}$. Closely related to these works are studies of the so-called ``lensing is low'' effect: when assuming the best-fit Planck2020 cosmology and fitting a model for galaxy clustering, the measured galaxy--galaxy lensing amplitude is overpredicted \citep{Leauthaud2017_MNRAS_467_3024}. It was shown that the measured lensing amplitude under Planck CMB parameters is around $15 - 35 \%$ lower than predicted, particularly on small scales\citep{Leauthaud2017_MNRAS_467_3024, Lange2021_MNRAS_502_2074, Amon2023_MNRAS_518_477} . Since the predicted lensing amplitude at fixed clustering is correlated with $S_8$ \citep{Yoo2006_ApJ_652_26}, these findings can also be seen as evidence of an $S_8$-tension. Furthermore, several recent studies \citep{Lange2022_MNRAS_509_1779, Chapman2022_MNRAS_516_617, Zhai2022_arXiv_2203_8999, Yuan2022_MNRAS_515_871} have applied a simulation-based modelling framework to the analysis of redshift-space clustering of galaxies. All four studies, using largely independent galaxy samples, find a preference for lower values for cosmic structure growth than the best-fit \cite{PlanckCollaboration2020_AA_641_6} $\Lambda$CDM model predicts.

For completeness, we point out that already in 2013 a combined full-scale analysis of projected galaxy clustering and galaxy-galaxy lensing based on SDSS data by \citet{Cacciato2013_MNRAS_430_767} yielded constraints on $\Omega_{\rm m}$ and $\sigma_8$ in excellent agreement with these more recent studies. However, at that point in time, those constraints where consistent with the then best-fit CMB constraints provided by the WMAP7 data \citep[][]{Komatsu2011_ApJS_192_18}, and thus did not signal any tension. Additionally, this study was based on an approximate analytical halo model and did not capture the effects of assembly bias. Similarly, \cite{Reid2014_MNRAS_444_476} performed a simulation-based full-scale analysis of redshift-space distortions in BOSS and also found a preference for a low structure growth amplitude. However, this study relied on re-scaling velocities in a single cosmological simulation which has been argued to lead to inaccurate results \citep{Zhai2019_ApJ_874_95}.

In this work, we built upon previous efforts by modelling the redshift-space clustering and galaxy--galaxy lensing of BOSS LOWZ galaxies. This work extends previous studies in several ways. Compared to the lensing studies of \cite{Wibking2020_MNRAS_492_2872} and \cite{Miyatake2022_PhRvD_106_3520}, we incorporate a model for galaxy assembly bias which has been shown to be important for modelling lensing on non-linear scales \citep{Lange2019_MNRAS_488_5771, Yuan2020_MNRAS_493_5551}. Furthermore, we measure and analyse updated high-precision galaxy--galaxy lensing measurements from the state-of-the-art DES Y3 and KiDS-1000 data sets with reduced systematic uncertainties. Compared to the redshift-space clustering-only analysis of \cite{Lange2022_MNRAS_509_1779}, the present work doubles the number of BOSS LOWZ galaxies, analysing nearly the entire BOSS LOWZ sample. Furthermore, we introduce and apply a new blinding (masking) methodology. Most importantly, the present study is the first { \it simulation-based full-scale and full-shape  joint cosmological analysis of galaxy-galaxy lensing and redshift-space clustering}.

Our paper is organised as follows. We start by introduction the observational data set and observabeles in section \ref{sec:observations}. Our modelling approach is described in section \ref{sec:modelling} and verified on mock catalogues in section \ref{sec:mocks}. The masking strategy is described and tested in section \ref{sec:masking}. In section \ref{sec:results}, we apply our analysis technique to observations and present the cosmological constraints. Finally, we discuss our results in section \ref{sec:discussion} and section \ref{sec:conclusions} presents our conclusions.

\section{Observations}
\label{sec:observations}

Here, we describe the observational data and the calculation of the summary statistics used to derive cosmological constraints.

\subsection{Observational data sets}

Our primary data set is the BOSS LOWZ galaxy sample. The clustering of BOSS LOWZ galaxies will be used as a summary statistic in our modelling. Additionally, we measure the galaxy--galaxy lensing effect around LOWZ galaxies using gravitational lensing catalogues from KiDS and DES.

\subsubsection{Baryon Oscillation Spectroscopic Survey}

Galaxies in BOSS LOWZ are targeted for spectroscopic observations if they fulfil a series of magnitude cuts aimed at selecting luminous red galaxies (LRGs) in the redshift range $0.15 \leq z \leq 0.5$. In the following, cuts on apparent magnitudes use cmodel magnitudes whereas colours are calculated using model magnitudes. The cuts are as follows:
\begin{eqnarray}
    r_{\rm cmod} &<& 13.5 + c_\parallel\, / 0.3 \label{eq:cut_c_parallel}\\
    |c_\perp| &<& 0.2 \label{eq:cut_c_perp}\\
    16 < &r_{\rm cmod}& < 19.6\, , \label{eq:m_cut}
\end{eqnarray}
where
\begin{equation}
    c_\parallel = 0.7 (g_{\rm mod} - r_{\rm mod}) + 1.2 (r_{\rm mod} - i_{\rm mod} - 0.18)
\end{equation}
and
\begin{equation}
    c_\perp = r_{\rm mod} - i_{\rm mod} - (g_{\rm mod} - r_{\rm mod}) / 4.0 - 0.18\, .
\end{equation}

\begin{table}
    \centering
    \begin{tabular}{c|c|c|c}
    Property & Sample A & Sample B & Sample C \\
    \hline
    $z_{\rm ref}$ & $0.25$ & $0.40$ & $0.40$ \\
    min. $z$ & $0.18$ & $0.30$ & $0.36$ \\
    max. $z$ & $0.30$ & $0.36$ & $0.43$ \\
    max. $M_r$ & $-20.412$ & $-20.558$ & $-21.208$ \\
    min. $c_\perp^0$ & $-0.216$ & $-0.154$ & $-0.166$ \\
    max. $c_\perp^0$ & $0.172$ & $0.234$ & $0.126$ \\
    max. $M_r - c_\parallel^0 / 0.3$ & $-25.874$ & $-26.729$ & $-26.901$ \\
    volume $V \, [\mathrm{Gpc}^3 \, h^{-3}]$ & $0.26$/$0.11$ & $0.22$/$0.10$ & $0.35$/$0.15$ \\
    $n_{\rm gal} \, [10^{-4} \, \mathrm{Mpc}^{-3} \, h^3]$ & $3.11$/$3.37$ & $3.13$/$3.14$ & $1.38$/$1.35$ \\
    \end{tabular}
    \caption{Definitions and properties of the samples analysed in this work. All three samples are designed to be subsets of the BOSS LOWZ sample that are roughly volume-limited in their respective redshift ranges. In the above table, $M_r$ is the absolute $r$-band magnitude and the superscript $^0$ indicates rest-frame colours. Both absolute magnitudes and colours are $k$-corrected to $z_{\rm ref}$. The final two rows indicate the volumes and galaxy number densities, split by NGC and SGC areas of the sky.}
    \label{tab:samples}
\end{table}

The above colour cuts, which are based on apparent photometric magnitudes, select galaxies that primarily fall in the redshift range, $0.15 \leq z \leq 0.5$. However, because the LOWZ target selection is based on apparent magnitudes, LOWZ galaxies are not uniformly selected in redshift. Particularly, galaxies of similar intrinsic luminosities and colours will have different apparent magnitudes depending on the redshift. However, such a selection would contradict our modelling approach, which implicitly assumes a volume-limited sample of red galaxies. Thus, additional selection cuts are needed to arrive at approximately volume-limited samples of galaxies. We follow \cite{Lange2022_MNRAS_509_1779} and construct three subsamples of the BOSS LOWZ galaxy sample in the redshift ranges, $0.18 < z \leq 0.30$, $0.30 < z \leq 0.36$ and $0.36 < z \leq 0.43$, whose cuts are based on absolute magnitudes and rest-frame colours, $k$-corrected to a reference redshift $z_{\rm ref}$. The samples and basic properties are described in Table~\ref{tab:samples}. Sample A in the Northern Galactic Cap (NGC) area is almost identical to the low-redshift sample in \cite{Lange2022_MNRAS_509_1779}. Conversely, samples B and C correspond to the single high-redshift sample in \cite{Lange2022_MNRAS_509_1779} but combined they have roughly $~60\%$ more galaxies per sky area. Additionally, compared to \cite{Lange2022_MNRAS_509_1779}, we also analyse galaxies from the Southern Galactic Cap (SGC) region. Taken together, these changes roughly double the sample size compared to \cite{Lange2022_MNRAS_509_1779} to roughly $3 \times 10^5$ galaxies altogether. Note that due to the slightly different photometric zero-points in the NGC and SGC regions, we opt to model and analyse galaxy samples from these two hemispheres separately. Finally, we point out that we do not use the so-called LOWZE2 and LOWZE3 samples in the NGC that have relied on an incorrect star–galaxy separation criterion \citep{Reid2016_MNRAS_455_1553}.

\subsubsection{Kilo Degree Survey}

We use galaxies imaged by KiDS as so-called source galaxies when measuring the galaxy--galaxy lensing effect. Specifically, we use the KiDS-1000 data set covering roughly $1000 \, \mathrm{deg}^2$ of the extra-galactic sky, roughly half of which overlaps with the BOSS survey footprint, with a source density $\sim 6 \, \mathrm{arcmin}^{-2}$ \citep{Giblin2021_AA_645_105}. Galaxies in KiDS-1000 are grouped into $5$ broad tomographic redshift bins. As described in \cite{Hildebrandt2021_AA_647_124}, the source redshift distribution $n(z)$ of each of the tomographic redshift bins has been derived using self-organizing maps. We use the tabulated $n(z)$ to convert gravitational tangential shears into estimates of the excess surface density $\Delta\Sigma$, as described in section \ref{subsec:lensing_measurements}.

\subsubsection{Dark Energy Survey}
In addition to KiDS-1000, we also use the DES Y3 weak lensing shape catalogues. DES Y3 covers $\sim 4000 \, \mathrm{deg}^2$, out of which roughly $800  \, \mathrm{deg}^2$ overlap with BOSS \citep{Amon2023_MNRAS_518_477}, with an effective source density of $6 \, \mathrm{arcmin}^{-2}$ \citep{Gatti2021_MNRAS_504_4312}. Similar to KiDS-1000, DES Y3 source galaxies are grouped into $4$ tomographic redshift bins. Source redshift distributions $n(z)$ are derived from a combination of self-organising maps, small-scale shear ratios, clustering redshifts, \citep{Myles2021_MNRAS_505_4249} and take into account blending effects in the photometry \citep{MacCrann2022_MNRAS_509_3371}.

\subsection{Galaxy clustering measurements}

Galaxy clustering is characterised by the two-point correlation function $\xi(s, \mu)$, which measures the excess probability of having a pair of galaxies separated by $s$, the three-dimensional separation and $\mu$, the cosine of the angle between $s$ and the line of sight. We use the Landy–-Szalay estimator \citep{Landy1993_ApJ_412_64} and correct for fibre collisions using the methodology presented in \cite{Guo2012_ApJ_756_127}.

Most of the information contained in the two-point correlation function can be described by its multipole moments,
\begin{equation}
    \xi_\ell (s) = \frac{2\ell + 1}{2} \int\limits_{-1}^{1} L_\ell (\mu) \xi (s, \mu) d \mu \, ,
    \label{eq:multipoles}
\end{equation}
where $L_\ell$ represents the Legendre polynomial of order $\ell$. We use $\ell = 0, 2$ and $4$, the so-called monopole, quadrupole, and hexadecapole moments of the redshift-space correlation function, respectively. The multipole moments contain information about galaxy peculiar velocities due to redshift-space distortions that change the apparent positions of galaxies along the line of sight. Thus, the redshift-space clustering of galaxies itself, even without gravitational lensing, contains information about cosmic structure growth. We also use the projected correlation function $\wp$
\begin{equation}
    \wp (\rp) = \int\limits_{-\pi_{\rm max}}^{+\pi_{\rm max}} \xi \left( s, \pi / s \right) d \pi \, ,
\end{equation}
where $\pi_{\rm max} = 80 \, \hmpc$ and $s = \sqrt{\pi^2 + \rp^2}$. The projected correlation function is nearly independent of galaxy peculiar velocities \citep{vandenBosch2013_MNRAS_430_725}. By combining the projected correlation function with the galaxy--galaxy lensing amplitude we can probe cosmological constraints that are practically independent of peculiar velocities.

Both $\xi(s)$ and $\wp(\rp)$ are measured in $14$ comoving logarithmic bins going from $0.1 \, \hmpc$ to $10^{1.8} \approx 63 \, \hmpc$. Uncertainties on the correlation function were derived from jackknife-resampling of $100$ roughly equal-area patches of the BOSS LOWZ sample, separately for the NGC and SGC areas. The cross-covariances between different clustering measures at different radial bins as well as different multipole moments of the redshift-space correlation function are taken into account. We apply the smoothing procedure described and tested in \cite{Lange2022_MNRAS_509_1779} in order to suppress noise in the resulting covariance matrix estimate.

\subsection{Gravitational lensing measurements}
\label{subsec:lensing_measurements}

In addition to the clustering properties of BOSS LOWZ galaxies, we also measure the galaxy--galaxy lensing effect around them. This is done by analysing the mean tangential ellipticities $e_t$ of source galaxies from KiDS and DES around BOSS LOWZ lens galaxies. The mean tangential ellipticity is related to the so-called excess surface density $\Delta\Sigma$ around lens galaxies, defined via
\begin{equation}
    \Delta\Sigma (\rp) = \langle \Sigma (<\rp) \rangle - \Sigma(\rp) \, ,
\end{equation}
where $\Sigma$ denotes the surface mass density, $\rp$ the projected distance in the frame of the lens galaxy, and the $\langle \Sigma (<\rp) \rangle$ is the mean surface density inside a circle of radius $\rp$ centred on the lenses. The induced tangential ellipticity depends on $\Delta\Sigma$ and $\Sigma_{\rm crit}$ defined as
\begin{equation}
    \Sigma_{\rm crit} (z_l, z_s) = \frac{c^2}{4\pi G} \frac{1}{(1 + z_l)^2} \frac{D_A(z_s)}{D_A(z_l) D_A(z_l, z_s)} \, ,
\end{equation}
where $z_l$ and $z_s$ are the redshifts of the lens and source galaxy, respectively, and $D_A$ denotes the angular diameter distance. In the weak lensing regime, $\Delta\Sigma \ll \Sigma_{\rm crit}$, gravitational lensing induces a tangential shear component,
\begin{equation}
    \gamma_t = \frac{\Delta\Sigma}{\Sigma_{\rm crit}} \, .
\end{equation}
Galaxies have intrinsic ellipticities, such that $e_t$ is a stochastic measure of $\gamma_t$ and one needs to stack a large number of lens--source pairs. We use the following estimator for the mean excess surface density of galaxies:
\begin{equation}
    \widehat{\Delta\Sigma} = \overline{\mathcal{M}}^{-1} \left[ \Delta\Sigma_l - \Delta\Sigma_r \right] \, .
\end{equation}
In the above equation, $\overline{\mathcal{M}}$ is an estimate of the mean multiplicative bias of the tangential ellipticity, $\Delta\Sigma_l$ is the uncorrected estimate for $\Delta\Sigma$ around lens galaxies and $\Delta\Sigma_r$ the analogous quantity around random points. In the absence of systematics, the excess surface density around random points is zero on average but subtracting this estimate also reduces the variance of the $\Delta\Sigma$ estimator \citep{Singh2017_MNRAS_471_3827}. The mean multiplicative shear bias differs slightly between KiDS and DES due to different shape measurement algorithms employed. For KiDS, our multiplicative shear bias estimate is
\begin{equation}
    \overline{\mathcal{M}}_{\rm KiDS} = 1 + \frac{\sum_{ls} w_{ls} m_s}{\sum w_{ls}} \, ,
\end{equation}
where $m$ is the multiplicative shear bias that depends on the source tomographic redshift bin \citep{Giblin2021_AA_645_105}, the sum $\sum_{ls}$ goes over all suitable lens--source pairs separated by $r_p$ and $w_{ls}$ is a weight associated with each galaxy pair. For DES, the shear bias correction factor is given by
\begin{equation}
    \overline{\mathcal{M}}_{\rm DES} = \left( 1 + \frac{\sum_{ls} w_{ls} m_s}{\sum w_{ls}} \right) \left( 1 + \frac{\sum_{ls} w_{ls} [R_T + R_{\rm sel}]}{\sum w_{ls}} \right) \, .
\end{equation}
In the above equation, $m$ is the multiplicative shear bias induced by blending \citep{MacCrann2022_MNRAS_509_3371}. Furthermore, $R_T$ and $R_{\rm sel}$ are the tangential component of the {\sc metacalibration} shear response and the selection response, respectively \citep{Huff2017_arXiv_1702_2600}. Finally, the raw $\Delta\Sigma$ estimator for lens galaxies is defined as
\begin{equation}
    \Delta\Sigma_l = \frac{\sum_{ls} w_{ls} e_t \widehat{\Sigma}_{{\rm crit}, ls}}{\sum_{ls} w_{ls}} \, ,
\end{equation}
where $\widehat{\Sigma}_{\rm crit}$ is an estimator of the intrinsic critical surface density of each lens--source pair. We do not know the precise redshift for each individual source galaxy and, instead, have estimates of the normalised source redshift distribution $n_{\hat{b}}(z)$\footnote{By default, the DES Y3 redshift distributions are not normalised in order to incorporate blending effects. In this work, we absorb the normalisation into the multiplicative shear bias $m$, instead.} in each tomographic source bin $\hat{b}$ \citep{Hildebrandt2021_AA_647_124, Myles2021_MNRAS_505_4249}. In order to obtain unbiased lensing amplitudes, we use
\begin{equation}
    \widehat{\Sigma}_{{\rm crit}, ls} = \left[ \int\limits_0^\infty \Sigma_{\rm crit}^{-1} (z_l, z_s) n_{\hat{b}}(z_s) d z_s \right]^{-1} \, ,
\end{equation}
where $n_{\hat{b}}(z_s)$ are the normalised intrinsic redshift distributions in each tomographic photometric redshift bin \citep{Hildebrandt2021_AA_647_124, Myles2021_MNRAS_505_4249}. Finally, the lens--source weights are designed to minimise shape noise,
\begin{equation}
    w_{ls} = \frac{w_s}{\widehat{\Sigma}_{{\rm crit}, ls}^{-2}} \, ,
\end{equation}
where $w_s$ is the source weight provided in the KiDS and DES shape catalogues. All lensing calculations are performed with the {\sc dsigma} galaxy--galaxy lensing package \citep{Lange2022_ascl_soft_4006} version 0.7.0. Note that we do not apply a lens weight $w_l$ and instead correct for fibre collisions using a nearest-neighbour correction \citep{Miyatake2015_ApJ_806_1}. We find that the alternative weighting by $w_l = w_{\rm noz} + w_{\rm cp} - 1$ \citep[see e.g.][]{Leauthaud2017_MNRAS_467_3024} results in a $\sim 1\%$ lower lensing signal than our default choice. Thus, the choice of fibre correction scheme does not significantly affect our conclusions.

We measure $\Delta\Sigma (\rp)$ in $14$ logarithmic bins going from $0.1$ to $\sim 63 \, \hmpc$. Covariance matrices for the lensing measurements are derived from jackknife re-sampling of $50$ roughly equal-area patches of overlap regions of BOSS LOWZ with KiDS and DES. For the BOSS NGC area, we only use the KiDS shape catalogue, i.e. ignoring the small overlap of BOSS and DES. For the SGC area, we only use the DES catalogues. We employ the same smoothing procedure as for the clustering measurements to suppress noise in our covariance matrix estimate. We do not take into account the cross-covariance between clustering and lensing measuremens since those are expected to be negligible \citep{Taylor2022_PhRvD_106_3536}, especially when considering that the overlap regions of BOSS with KiDS and DES constitute only a small part of the total BOSS footprint. Finally, we neglect systematic uncertainties in the lensing measurements stemming from photometric redshift calibration and multiplicative shear bias corrections. These uncertainties are at most $1.5\%$ \citep{Amon2023_MNRAS_518_477} and would translate into a $\sim 1\%$ systematic uncertainty on $S_8$, significantly below statistical uncertainties presented in section \ref{sec:results}.

\section{Modelling}
\label{sec:modelling}

In this section, we describe our simulation-based modelling framework which closely follows the one presented in \cite{Lange2022_MNRAS_509_1779}. Thus, we only describe the most salient points here and refer the reader to \cite{Lange2022_MNRAS_509_1779} for a more in-depth discussion.

\subsection{Cosmological simulations}

We use the {\sc Aemulus} cosmological simulations \citep{DeRose2019_ApJ_875_69} to make predictions for galaxy clustering and galaxy--galaxy lensing. {\sc Aemulus} is a suite of $40$ simulations with different cosmological parameters. Each simulation traces structure formation in a $w$CDM Universe using $(1400)^3$ particles in a $(1050 \, \hmpc)^3$ cubic volume. As discussed in \cite{DeRose2019_ApJ_875_69}, the resolution of these simulations is sufficient to be used in the study of non-linear clustering of LRGs. Dark matter haloes in the simulations are identified with the {\sc Rockstar} halo finder \citep{Behroozi2013_ApJ_762_109}. In the following, we will only use parent haloes, i.e. no subhaloes, with a mass $M$ at or above $100$ times the particle mass $m_p$, where $m_p = 3.51 \times 10^{10} ( \Omega_{\rm m} / 0.3 ) h^{-1} M_\odot$.

\subsection{Galaxy--halo connection model}

We use a Halo Occupation Distribution \citep[HOD; ][]{Berlind2002_ApJ_575_587, Bullock2002_MNRAS_329_246, Zheng2007_ApJ_667_760} model as the basis for our galaxy--halo connection. In particular, we parameterise the average number of central and satellite galaxies expected in a halo as a function of its mass $M$ and maximum circular velocity $V_{\rm max}$. The average number of central galaxies as a function of halo mass $M$ is given by
\begin{equation}
    \langle N_{\rm cen} | M \rangle = \frac{f_\Gamma}{2} \left( 1 + \mathrm{erf} \left[ \frac{\log M - \log M_{\rm min}}{\sigma_{\log M}} \right] \right),
\end{equation}
where $f_\Gamma$, $\log M_{\rm min}$ and $\sigma_{\log M}$ are free parameters. The parameter $f_\Gamma$ models incompleteness in the selection of LRGs \citep{Leauthaud2016_MNRAS_457_4021}. The average number of satellites is given by
\begin{equation}
    \langle N_{\rm sat} | M \rangle = \left( \frac{M - M_0}{M_1} \right)^\alpha
\end{equation}
with $M_0$, $M_1$, and $\alpha$ being free parameters. The dependence on $V_{\rm max}$ is modelled via the decorated HOD framework \citep{Hearin2016_MNRAS_460_2552}, a natural extension to the standard, mass-only HOD approach. This is necessary to model the impact of galaxy assembly bias \citep{Gao2005_MNRAS_363_66, Wechsler2006_ApJ_652_71, Zentner2014_MNRAS_443_3044} and its degeneracy with cosmological parameters \citep{Lange2019_MNRAS_488_5771, Lange2019_MNRAS_490_1870, Yuan2019_MNRAS_486_708}. In short, we perturb the average number of expected galaxies in a halo of mass $M$ based on whether $V_{\rm max}$ is above or below average of haloes at that mass. The perturbation is described by
\begin{equation}
    \langle N_{\rm cen} | M, V_{\rm max} \rangle = \langle N_{\rm cen} | M \rangle \pm A_{\rm cen} \left( \frac{1}{2} - \left| \frac{1}{2} - \langle N_{\rm cen} | M \rangle \right| \right),
    \label{eq:dHOD}
\end{equation}
for centrals and
\begin{equation}
    \langle N_{\rm sat} | M, V_{\rm max} \rangle = (1 \pm A_{\rm sat}) \langle N_{\rm sat} | M \rangle,
\end{equation}
for satellites. In both cases, $\pm$ indicates $+$ when $V_{\rm max}$ is above the median at that mass and $-$ otherwise. This adds two more free parameters, $A_{\rm cen}$ and $A_{\rm sat}$, varying in the range $[-1, +1]$. Once average galaxy numbers are specified, the distribution of galaxy numbers follow Bernoulli and Poisson distributions for centrals and satellites, respectively. We note that several recent theoretical and observational studies indicate the possibility of non-Poisson satellite number distributions \citep[see, e.g.,][]{Dvornik2022_arXiv_2210_3110, ChavesMontero2022_arXiv_2211_1744}. In \cite{Lange2022_MNRAS_509_1779}, we showed that the assumption of a non-Poisson satellite distribution did not significantly affect cosmological constraints from redshift-space clustering. Similarly, non-Poisson numbers are not expected to have a substantial impact on the galaxy--galaxy lensing predictions in the two-halo regime that we are modelling \citep[see, e.g.,][]{Zu2020_arXiv_2010_1143, Lange2022_MNRAS_509_1779, ChavesMontero2022_arXiv_2211_1744}.

Centrals and satellites have different phase-space coordinates. Central galaxies coincide spatially with the halo centre but have additional random Gaussian velocities along the line of sight \citep{Reid2014_MNRAS_444_476, Guo2015_MNRAS_446_578, Guo2015_MNRAS_453_4368} with scatter $\sigma$,
\begin{equation}
    \sigma = \frac{\alpha_{\rm c} V_{\rm vir}}{\sqrt{3}}.
\end{equation}
This simulates the velocities of central galaxies with respect to the halo centre due to the unrelaxed state of the halo \citep{Ye2017_ApJ_841_45}. The free parameter $\alpha_{\rm c}$ is commonly known as the central velocity bias parameter. Satellite galaxies follow a Navarro--Frenk--White \citep[NFW; ][]{Navarro1997_ApJ_490_493} profile with respect to the halo centre,
\begin{equation}
    n(r) \propto \frac{1}{r / r_s \left( 1 + r / r_s \right)^2} \, .
\end{equation}
The concentration parameter $c_{\rm sat} = r_s / r_{\rm h}$, where $r_{\rm h}$ is the host halo radius, is allowed to be different than that of the dark matter distribution in the same halo, $c_{\rm dm}$,
\begin{equation}
    c_{\rm sat} = \eta c_{\rm dm} \, .
\end{equation}
Similar to central galaxies, satellites follow the halo velocity on average. We add additional velocities with respect to the halo by drawing from a Gaussian distribution with scatter derived from the spherically symmetric, anisotropy-free Jeans equation. The derived satellite velocities are likely an approximation since satellite populations will not be perfectly spherically symmetric, without velocity anisotropy or dynamically relaxed. Thus, we apply an additional free scaling factor $\alpha_{\rm s}$ (the satellite velocity bias parameter) to the velocity scatter derived from the Jeans equation. Overall, the phase-space coordinates of galaxies add an additional three free parameters, $\alpha_{\rm c}$, $\alpha_{\rm s}$, and $\eta$. Ultimately, our model for galaxies has $11$ free parameters, which we list in Table~\ref{tab:galaxy-halo_connection} for reference, that we allow to vary when fitting for cosmology.

\begin{table}
    \centering
    \begin{tabular}{l|l|l}
    Parameter & Description & Range \\
    \hline
    $\log M_{\rm min}$ & low-mass cut-off for $\langle N_{\rm cen} \rangle$ & $[12.5, 14.0]$ \\
    $\sigma_{\log M}$ & low-mass transition for $\langle N_{\rm cen} \rangle$ & $[0.1, 1.0]$ \\
    $f_\Gamma$ & incompleteness for $\langle N_{\rm cen} \rangle$ & $[0.5, 1.0]$ \\
    $\log M_0$ & low-mass cut-off for $\langle N_{\rm sat} \rangle$ & $[12.0, 15.0]$ \\
    $\log M_1$ & characteristic halo mass for $\langle N_{\rm sat} \rangle$ & $[13.5, 15.0]$ \\
    $\alpha$ & power-law index for $\langle N_{\rm sat} \rangle$ & $[0.5, 2.0]$ \\
    $A_{\rm cen}$ & central galaxy assembly bias & $[-1.0, 1.0]$ \\
    $A_{\rm sat}$ & satellite galaxy assembly bias & $[-1.0, 1.0]$ \\
    $\log \eta$ & satellite spatial bias & $[-0.5, 0.5]$ \\
    $\alpha_{\rm c}$ & central velocity bias & $[0.0, 0.4]$ \\
    $\alpha_{\rm s}$ & satellite velocity bias & $[0.8, 1.2]$ \\
    \end{tabular}
    \caption{All galaxy--halo connection parameters modelled in this work together with a short description and flat prior ranges used for fitting.}
    \label{tab:galaxy-halo_connection}
\end{table}

\subsection{Data likelihood}

We use {\sc Halotools} \citep{Hearin2017_AJ_154_190} to compute galaxy clustering and lensing observables for a given simulation with cosmological parameters $\mathcal{C}$ and galaxy--halo connection parameters $\mathcal{G}$. For sample $A$, we model the observables using simulation outputs at redshift $0.25$ whereas for samples $B$ and $C$, we use $z=0.40$ snapshots\footnote{For sample B, there is an apparent mismatch between the mean redshift of the sample, $z=0.33$, and the redshift of the simulation output used to analyse it, $z=0.40$. As shown in \cite{Lange2022_MNRAS_509_1779}, this should have a negligible impact on studies using RSDs since these observations are primarily sensitive to $f(z) \sigma_8 (z)$, which evolves very slowly with redshift. However, the predicted lensing signal at fixed clustering roughly scales as $\sigma_8(z)$ \citep{Yoo2006_ApJ_652_26}. Thus, to account for the redshift mismatch, we multiply lensing predictions for sample B by $\sigma_8(0.33) / \sigma_8(0.40) = 1.04$.}. We make use of the distant observer approximation and in all cases project galaxy populations onto each of the three simulation axes. When making predictions for galaxy clustering, we take into account the Alcock--Paczynski effect \citep{Alcock1979_Natur_281_358} by correcting simulation coordinates for the assumed cosmology when obtaining the clustering measurements in section \ref{sec:observations}. Similarly, we correct the $\Delta\Sigma$ predictions for the assumed cosmology following the methodology described in \cite{More2013_ApJ_777_26} and approximating $\Delta\Sigma \propto r_p^{-1}$. To speed up the calculation of the clustering and lensing predictions for a given {\sc Aemulus} simulation, we make use of a tabulation method for galaxy correlation functions \citep{Zheng2016_MNRAS_458_4015} as implemented in {\sc TabCorr}\footnote{\url{https://github.com/johannesulf/TabCorr}} version 1.0.0. For a given clustering and lensing prediction, the likelihood $\mathcal{L}$ of the observed data vector $\mathbf{D}$ is computed as
\begin{equation}
    \begin{split}
        \ln \mathcal{L} (\mathbf{D} | \mathcal{C}, \mathcal{G}) = \frac{1}{2} &\left[ (n_{\rm gal} - \hat{n}_{\rm gal})^2 \sigma_{n_{\rm gal}}^{-2} + (\xi - \hat{\xi})^{\rm T} \Sigma_\xi^{-1 } (\xi - \hat{\xi}) \right. \\
        &\left. + (\Delta\Sigma - \widehat{\Delta\Sigma})^{\rm T} \Sigma_{\Delta\Sigma}^{-1} (\Delta\Sigma - \widehat{\Delta\Sigma}) \right] \, ,
    \end{split}
    \label{eq:likelihood}
\end{equation}
where $\xi$ denotes the combination of all galaxy clustering measurements, e.g. $\xi_0$, $\xi_2$ and $\xi_4$ or just $\wp$.

We use three different choices of data sets $\mathbf{D}$. The first set, which we call ``RSD-only'' consists of the three redshift-space clustering multipole moments, $\xi_0$, $\xi_2$, and $\xi_4$, and no lensing measurements. The second set labelled ``$w_p + \Delta\Sigma$'' combines the projected correlation function $\wp$ with galaxy--galaxy lensing $\Delta\Sigma$. Finally, the set called ``RSD + $\Delta\Sigma$'' consists of the redshift-space clustering multipole moments and the lensing amplitude. All clustering measurements, $\xi_0$, $\xi_2$, $\xi_4$, and $\wp$ are fitted on scales larger than $0.4 \hmpc$, as discussed and motivated in \cite{Lange2022_MNRAS_509_1779}. Contrary, for lensing, we only consider scales $2.5 \, \hmpc < \rp < 25 \hmpc$. The lower limit is chosen to avoid biases associated with baryonic feedback \citep{Leauthaud2017_MNRAS_467_3024, Lange2019_MNRAS_488_5771, Amodeo2021_PhRvD_103_3514} which we do not model in this analysis. The upper limit is chosen in order to avoid biases in the covariance matrix estimate related to the size of the jackknife fields \citep{Shirasaki2017_MNRAS_470_3476}.

We note that for all three choices of two-point correlation functions, we use the number density of galaxies, $n_{\rm gal}$, as a constraint. This observable tightly limits one dimension of the HOD parameter space due to the requirement that total number density of galaxies matched the observed one. At the same time, the galaxy--halo connection model has significant flexibility to change the predicted number density without affecting the clustering predictions. For example, the predictions for the two-point correlation functions are unaffected if the number of galaxies in all halos was changed by constant factor. Consequently, if we replace $f_\Gamma \rightarrow x \, f_\Gamma$ and $M_1 \rightarrow x^{-1/\alpha} \, M_1$, the number density prediction changes to $\hat{n}_{\rm gal} \rightarrow x \, \hat{n}_{\rm gal}$ while leaving the clustering and lensing predictions unaffected\footnote{This is strictly true only for $A_{\rm cen} = 0$. However, observations do not tightly constrain $A_{\rm cen}$ and it is always consistent with $0$.}. Ultimately, we do not expect that the number density constraint significantly affects the cosmology result since neither $f_\Gamma$ nor $\log M_1$ are tightly constrained by the observational data or pile up against the prior ranges, i.e. $f_\Gamma \leq 1$.

Equation (\ref{eq:likelihood}) describes the likelihood of any simulation and its underlying cosmology $\mathcal{C}$ as a function of galaxy parameters $\mathcal{G}$. However, ultimately, we want to obtain a constraint on $\mathcal{C}$ alone for which we need $\mathcal{L} (\mathbf{D} | \mathcal{C})$. Therefore, we first have to marginalise the data likelihood over the galaxy parameters $\mathcal{G}$. In \cite{Lange2019_MNRAS_490_1870} and \cite{Lange2022_MNRAS_509_1779}, $\mathcal{L} (\mathbf{D} | \mathcal{C})$ is the evidence $\mathcal{Z} (\mathbf{D} | \mathcal{C})$,
\begin{equation}
    \mathcal{Z} (\mathbf{D} | \mathcal{C}) = \int \mathcal{L} (\mathbf{D} | \mathcal{C}, \mathcal{G}) P(\mathcal{G}) d\mathcal{G}\, .
    \label{eq:evidence}
\end{equation}
As an alternative, one can also use the profile likelihood,
\begin{equation}
    \mathcal{L}_p (\mathbf{D} | \mathcal{C}) = \max_{\mathcal{G}} \mathcal{L} (\mathbf{D} | \mathcal{C}, \mathcal{G}) \, ,
    \label{eq:profile_likelihood}
\end{equation}
i.e. the maximum likelihood obtained over all galaxy--halo connection parameters, as a replacement for the evidence. The advantage of this approach is that $\mathcal{L} (\mathbf{D} | \mathcal{C})$ becomes independent of poorly motivated priors on the galaxy model. We will compare the performance of both approaches in section \ref{sec:mocks}. To calculate the evidence, maximum likelihood and posteriors on galaxy--halo connection parameters for each simulation, we employ the {\sc nautilus}\footnote{\url{https://github.com/johannesulf/Nautilus}} sampler version 0.2.1 (Lange, in prep.). We use $3000$ live points, discard points during the exploration phase and run the sampler until an effective sample size of $100,000$ is achieved.

\subsection{Cosmological inference}
\label{subsec:cosmology_fitting}

Once we have computed the summary statistic, either the evidence $\mathcal{Z}$ or the profile likelihood $\mathcal{L}_p$, we have to characterise the dependence of those summary statistics on the cosmology of each simulation. Full-scale redshift-space clustering is sensitive to $f(z) \sigma_8(z)$ \citep{Lange2022_MNRAS_509_1779}, where $f$ is the linear growth rate. Conversely, projected galaxy clustering and galaxy--galaxy lensing are sensitive to both $\Omega_{\rm m}$ and $\sigma_8(z)$ \citep{Yoo2006_ApJ_652_26}. Since our analysis exhibits {\it joint} dependence upon these three cosmological parameters, we elect to model the summary statistic also as a function of three parameters that fully specify $f(z)$, $\sigma_8(z)$, and $\Omega_{\rm m}$. Here, we choose these three cosmological parameters to be $S_8$, $\Omega_{\rm m}$, and $w$, the Dark Energy equation of state. We assume that the dependence of $\mathcal{Z}$ and $\mathcal{L}_p$ on $S_8$, $\Omega_{\rm m}$, and $w$ can be modelled as a multi-variate skew-normal distribution \citep{Azzalini1996}. For reference, in one dimension, the probability distribution function (PDF) of a skew-normal distribution is given by,
\begin{equation}
    f(x) = \frac{2}{\sigma} \phi \left( \frac{x - \mu}{\sigma} \right) \Phi \left( \lambda \frac{x - \mu}{\sigma} \right) \, ,
\end{equation}
where $\phi$ and $\Phi$ are the PDF and cumulative distribution function (CDF) of the standard normal, respectively. This functional form is motivated empirically \citep{Lange2019_MNRAS_490_1870} and is a natural extension of the assumption of a Gaussian posterior. A three-dimensional skew-normal has $12$ free parameters: three means $\mu$, three standard deviations $\hat{\sigma} = \sigma / (1 + \sigma) \in [0, 1]$, three skew parameters $\delta = \lambda / \sqrt{1 + \lambda^2} \in [-1, +1]$ as well as three parameters for the off-diagonals of the covariance matrix, i.e. $r_{S_8, \Omega_{\rm m}}$, $r_{S_8, w}$, and $r_{\Omega_{\rm m}, w} \in [-1, +1]$.

We obtain our full posterior constraints on $S_8$, $\Omega_{\rm m}$, and $w$ as follows. Beginning with the collection of $40$\footnote{In practice, we always exclude the simulation Box023 since it has an unusually low $S_8 = 0.595$, well below the second lowest value of $0.703$. The motivation is that this simulation is almost always confidently excluded by observations and we do not want its very low likelihood to influence the fit to $\mathcal{L} (\mathbf{D} | \mathcal{C})$ in regions of cosmological parameter space allowed by observations.} values for $\mathcal{Z} (\mathbf{D} | \mathcal{C})$ or $\mathcal{L}_p (\mathbf{D} | \mathcal{C})$ (depending on which we use for the summary statistic), we approximate the distribution of these values with a skew-normal distribution and use an MCMC to derive full posterior distributions on the $12$ skew-normal hyper-parameters plus one additional free parameter for the likelihood scatter of each simulation. Each point in this 13-dimensional hyper-parameter space represents a particular skew-normal approximation to the cosmology-dependence of our summary statistic. We draw $N$ samples of our hyper-parameters based on the posteriors on these quantities, and we derive our constraints on $S_8$, $\Omega_{\rm m}$ and $w$, i.e. $P (S_8, \Omega_{\rm m}, w)$ based on the superposition of the resulting collection of normalised skew-normals. As in \cite{Lange2022_MNRAS_509_1779}, we place explicit flat priors on cosmological parameters, in this case $S_8$, $\Omega_{\rm m}$ and $w$. This is done to take into account that we cannot reliably extrapolate the dependence of $\mathcal{Z}$ and $\mathcal{L}_p$ on parts of the cosmological parameter space not probed by the {\sc Aemulus} simulations. The prior is determined by placing a three-dimensional minimum-volume enclosing ellipsoid around the parameter combinations of the {\sc Aemulus} simulations. Combinations of $S_8$, $\Omega_{\rm m}$, and $w$ outside this ellipse are assigned $0$ prior probability. We refer the reader to \cite{Lange2022_MNRAS_509_1779} for a detailed discussion and motivation of the fitting procedure described here. In order to verify that our conclusions are robust to our assumption that the cosmology-dependence of $\mathcal{Z}$ and $\mathcal{L}_p$ is well-described by a skew-normal form, in appendix \ref{sec:gp_modelling} we conduct an alternate analysis in which we alternatively use a Gaussian Process regression to approximate these distributions, finding very similar results. Finally, we note that our final constraints on cosmology should not be regarded as being model-independent or valid outside the $w$CDM cosmological parameter space probed by the {\sc Aemulus} simulations.

\section{Verification on mock catalogues}
\label{sec:mocks}

Before applying our analysis method to the observational data described in section \ref{sec:observations}, we test our methods on mock catalogues to ensure that our cosmological inferences do not suffer from significant biases.

\subsection{Mock observations}

The mock catalogues used here are described in more detail in \cite{Lange2022_MNRAS_509_1779}. They are created from the UNIT simulations \citep{Chuang2019_MNRAS_487_48} and the associated {\sc Rockstar} halo catalogues. We populate haloes in the $z=0.25$ outputs of each of the four UNIT simulations with galaxies using the subhalo abundance matching framework \citep[SHAM; ][]{Vale2004_MNRAS_353_189, Conroy2006_ApJ_647_201}. In essence, we place galaxies at the centres of the haloes with the highest $\alpha \log V_{\rm peak} + (1 - \alpha) \log V_{\rm vir}$, where $V_{\rm vir}$ is the halo virial velocity at the epoch of peak halo mass, $V_{\rm peak}$ the corresponding value of $V_{\rm max}$ at that epoch and $\alpha = 0.73$ \citep{Lehmann2017_ApJ_834_37}. We populate haloes until the number density of galaxies matches that of the LOWZ $0.18 \leq z \leq 0.30$ sample. Haloes in this procedure include both parent haloes as well as subhaloes such that satellite galaxies are naturally accounted for. As described in \cite{Lange2022_MNRAS_509_1779}, small modifications to the SHAM procedure are applied to account for scatter between galaxy and halo properties. Once mock galaxy catalogues are created, we compute mock observables assuming a distant observer approximation. Since we have four simulations of $1 \, \mathrm{Gpc}^3 \, h^{-3}$ volume each, and we consider all three simulation axes as the line of sight, the mock measurements have an effective volume of $\sim 12 \, \mathrm{Gpc}^3 \, h^{-3}$ \citep{Smith2021_MNRAS_500_259}.

\begin{figure}
    \centering
    \includegraphics{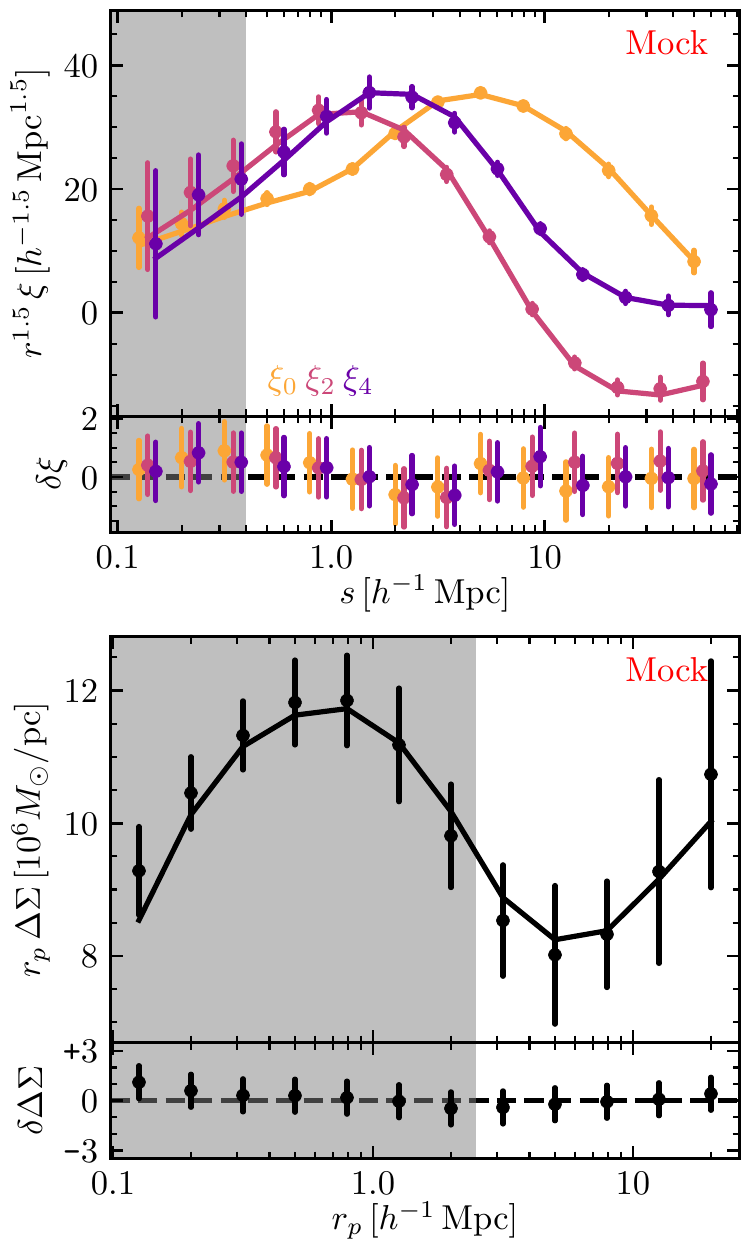}
    \caption{Mock observations of galaxy clustering in redshift space (top) and galaxy--galaxy lensing (bottom). Dots indicate the mock measurements themselves, error bars the assumed observational uncertainty, and the solid line is the best-fit model prediction. The lower panels show the difference between mock observations and best-fit model predictions in units of the observational uncertainty. Grey backgrounds indicate the ranges of the small-scale data that are not included in the fit.}
    \label{fig:mock_fit}
\end{figure}

The mock observations for galaxy clustering and galaxy--galaxy lensing are displayed in Fig. \ref{fig:mock_fit}. Error bars represent the observational uncertainties of the $0.18 \leq z \leq 0.30$ NGC (SGC) sample for galaxy clustering (galaxy--galaxy lensing). In the following, we will use these same observational uncertainties, i.e. covariance matrix, to fit our model to the mock data set. The best-fit model predictions, marginalised over all galaxy parameters and all $40$ simulations, are shown in Fig. \ref{fig:mock_fit} by the solid lines. Overall, our HOD-based galaxy model, which is an empirical model that is founded upon different assumptions than the SHAM model used to produce the mock catalogues, can produce good fits to the mock data.

\subsection{Cosmology results}

\begin{figure}
    \includegraphics{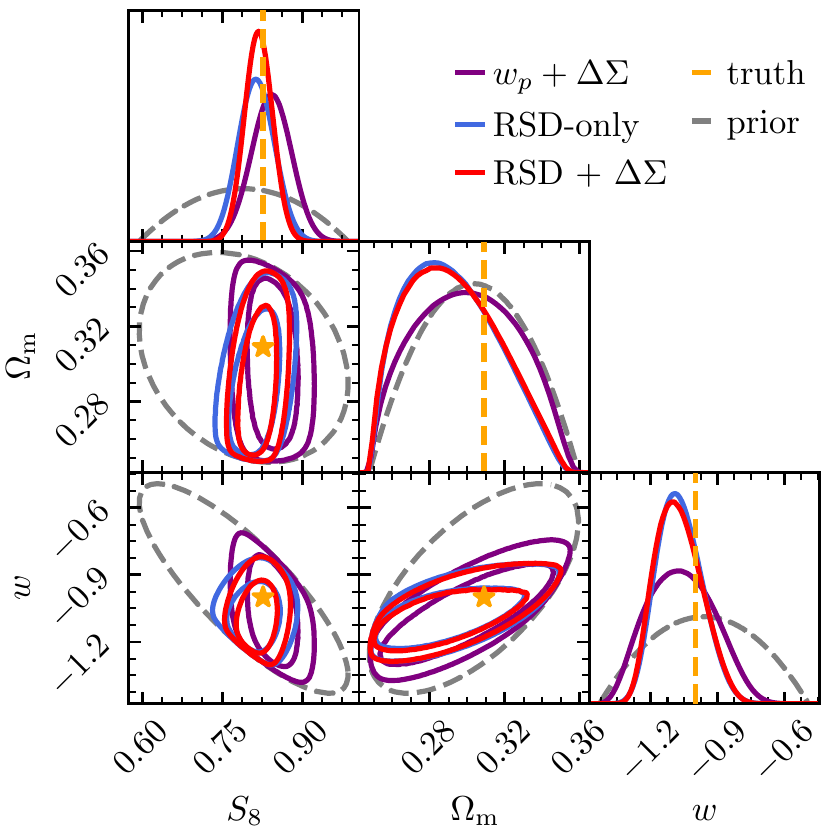}
    \caption{Posterior constraints on cosmological parameters when analysing the mock data set derived from the UNIT simulations. We show the results from analysis of projected clustering and galaxy--galaxy lensing (purple), the study of redshift-space multipoles (blue), and the combination of the redshift-space clustering and galaxy--galaxy lensing (red). In all panels we highlight the cosmological parameters of the UNIT simulations (yellow) we seek to recover. Finally, the grey lines indicate our prior: in the off-diagonal panels it shows the range and in the diagonal elements the implicit one-dimensional prior implied by projecting the volume of a three-dimensional ellipsoid onto one axis. For this figure, the evidence $\mathcal{Z}$ was used as the summary statistic for each of the $40$ {\sc Aemulus} simulations.}
    \label{fig:posterior_m_z}
\end{figure}

\begin{figure}
    \includegraphics{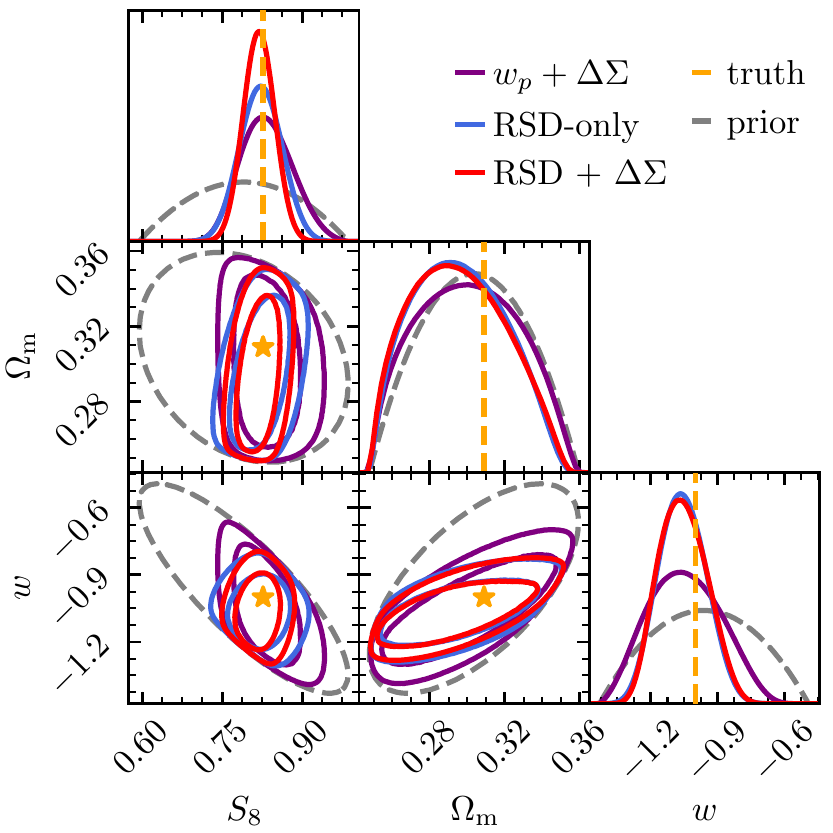}
    \caption{Similar to Fig. \ref{fig:posterior_m_z} but with the profile likelihood $\mathcal{L}_p$ instead of the evidence used as the summary statistic.}
    \label{fig:posterior_m_l}
\end{figure}

Cosmological parameter constraints for the simulated dataset are shown in Figs. \ref{fig:posterior_m_z} and \ref{fig:posterior_m_l}. In each of the figures, red lines indicate posterior constraints when fitting both RSDs and lensing, i.e. $\xi_0$, $\xi_2$, $\xi_4$ and $\Delta\Sigma$. Blue lines indicate posterior constraints when only redshift-space clustering is considered in the fit, and purple lines indicate results obtained without redshift-space clustering. Since galaxy--galaxy lensing is only sensitive to cosmology in combination with galaxy clustering \citep{Yoo2006_ApJ_652_26}, we also include the projected two-point correlation function, $w_p$, in that particular fit. Since $w_p$ is obtained by projecting the redshift-space two-point correlation function along the line of sight, it is mostly insensitive to redshift-space distortions. As shown in \cite{Lange2022_MNRAS_509_1779}, $w_p$ by itself has little cosmological information.

The difference between Figs. \ref{fig:posterior_m_z} and \ref{fig:posterior_m_l} is that the former uses the evidence $\mathcal{Z}$ as the summary statistic, whereas the latter uses the profile likelihood $\mathcal{L}_p$. We see that our analysis places strong constraints on $S_8$, as expected. Both redshift-space clustering and galaxy--galaxy lensing obtain roughly comparable constraints on $S_8$ with slightly different degeneracies with respect to the other cosmological parameters. On the other hand, neither RSDs nor lensing lead to strong constraints on $\Omega_{\rm m}$, i.e. the posterior PDF is very similar to the prior. Finally, we nominally get noteworthy constraints on $w$. However, a significant fraction of this constraint may originate indirectly from the constraint on $S_8$. Specifically, our assumed prior implies a strong correlation between $S_8$ and $w$ such that any constraint on $S_8$ also leads to a strong constraint on $w$, as is evident from the lower left panels in each of the two figures.

We find that both approaches, either employing the evidence or the profile likelihood, are successful in recovering the input cosmology of the UNIT simulations. Additionally, the fit using the evidence as the summary statistic produces quantitatively  similar results to the analysis using the profile likelihood. As described before, the effective volume of the mock observations is roughly $\sim 12 \, \mathrm{Gpc}^3 \, h^{-3}$, a factor of $\sim 40$ larger than the observational volume from which the assumed observational uncertainties come from. Thus, we expect our results to reproduce the input to much better than the assumed observational uncertainty. With this in mind, it is noteworthy that the posterior constraints on $S_8$ are slightly off-centred when fitting the evidence for either the case of redshift-space clustering or galaxy--galaxy lensing in isolation; this is not the case for the fit involving the profile likelihood. Moreover, the profile likelihood has a comparatively stronger theoretical motivation since it is independent of the arbitrary priors on the galaxy--halo connection. For these reasons, we will choose the profile likelihood as the default summary statistic when analysing the real data in section~\ref{sec:results}.

\section{Masking strategy}
\label{sec:masking}

For scientific studies, we want to protect the experiment and analysis design from confirmation bias of the scientists leading the analysis. A common strategy to avoid this issue is to ``blind'' the analysis with respect to the key scientific results while still being able to perform critical null tests. In our example, we wish to make analysis choices insensitive to the cosmological constraints, i.e. $S_8$, while still being able to judge, for example, the goodness of fit of our model. Throughout this work, we will use the term ``masking'' instead of the more commonly used term ``blinding''.

\subsection{Proposed method}

A masking procedure can in principle happen at various stages of the analysis. In the case of posterior masking, one would simply hide or randomly perturb the final cosmological posterior. Data vector masking involves perturbing the summary statistics, i.e. $\xi$ and $\Delta\Sigma$, in ways that also randomly perturb the final cosmological result. Finally, masking can be implemented at the catalogue level such that both summary statistics and cosmological posterior are affected. Data vector and catalogue level masking are harder to implement than posterior masking but are more robust in the sense that they are less prone to accidental unmasking. We refer the reader to \cite{Muir2020_MNRAS_494_4454} for a detailed discussion and motivation behind different approaches. In this work, we will apply a data vector masking procedure.

The method employed here is a variation of the data vector masking procedure described in \cite{Muir2020_MNRAS_494_4454}. Let us assume that we have an unperturbed data vector $\mathbf{D}$ and a model for that data vector $\widehat{\mathbf{D}} (\mathbf{\theta})$ where $\theta$ represents the model parameters we wish to constrain with the analysis. \cite{Muir2020_MNRAS_494_4454} propose perturbing $\mathbf{D}$ by
\begin{equation}
    \Delta \mathbf{D} (\Delta \mathbf{\theta}) = \widehat{\mathbf{D}} (\mathbf{\theta}_{\rm fid} + \Delta \mathbf{\theta}) - \widehat{\mathbf{D}} (\mathbf{\theta}_{\rm fid}) \, ,
\end{equation}
where $\Delta \mathbf{\theta}$ is an offset in the model and $\mathbf{\theta}_{\rm fid}$ are suitably chosen set of fiducial model parameters. Under idealised conditions, adding $\Delta \mathbf{D} (\Delta \mathbf{\theta})$ to the unperturbed data vector will shift the posterior of $\theta$ by $\Delta \mathbf{\theta}$ while leaving the goodness of fit, i.e. the minimum $\chi^2$, unchanged. In practice, these idealised conditions are not perfectly met, such that the above statements are only approximately true and the masking procedure needs to be validated with simulations first \citep{Muir2020_MNRAS_494_4454}.

In our analysis, we wish to mask the final constraints on $S_8$. Constraints on the galaxy--halo connection are of less interest in this study and could be described as nuisance parameters. The original method proposed in \cite{Muir2020_MNRAS_494_4454} would calculate $\Delta \mathbf{D} (\Delta \mathbf{\theta})$ by looking at the difference in predictions for two cosmologies with different $S_8$ values while keeping the galaxy--halo connection parameters fixed. We may choose to perturb $S_8$ by up to $\Delta S_8 = 0.1$, which, as we will show later, would correspond to up to $4 \sigma$ shifts in the final $S_8$ posterior. However, the impact of cosmology and galaxy--halo connection parameters on the data vector is often degenerate. For example, the large-scale clustering amplitude is sensitive to $b \sigma_8$ where $b$ is the galaxy bias. As a result, implementing the method described in \cite{Muir2020_MNRAS_494_4454} could result in data vector shifts that are large with respect to the observational uncertainties, i.e. significantly larger than $4 \sigma$.

Here, we propose a slight variation of the method described in \cite{Muir2020_MNRAS_494_4454}. When calculating $\Delta \mathbf{D}$, instead of only changing $S_8$ while keeping all other model parameters fixed we instead change $S_8$ and then vary all other model parameters such that the difference between $\widehat{\mathbf{D}} (\mathbf{\theta}_{\rm fid} + \Delta \mathbf{\theta})$ and $\widehat{\mathbf{D}} (\mathbf{\theta}_{\rm fid})$ is minimised. Here, we define the difference between the data vectors as their $\chi^2$ difference.  This ensures that the shift in the data vector is as small as possible while ensuring the desired shift in $S_8$. For example, a $4 \sigma$ shift in $S_8$ would result in a roughly $4 \sigma$ significant shift of the data vector. Overall, this procedure seems more likely than the original \citet{Muir2020_MNRAS_494_4454} method to preserve the goodness of fit after masking.

\subsection{Specific implementation}

\begin{figure}
    \includegraphics{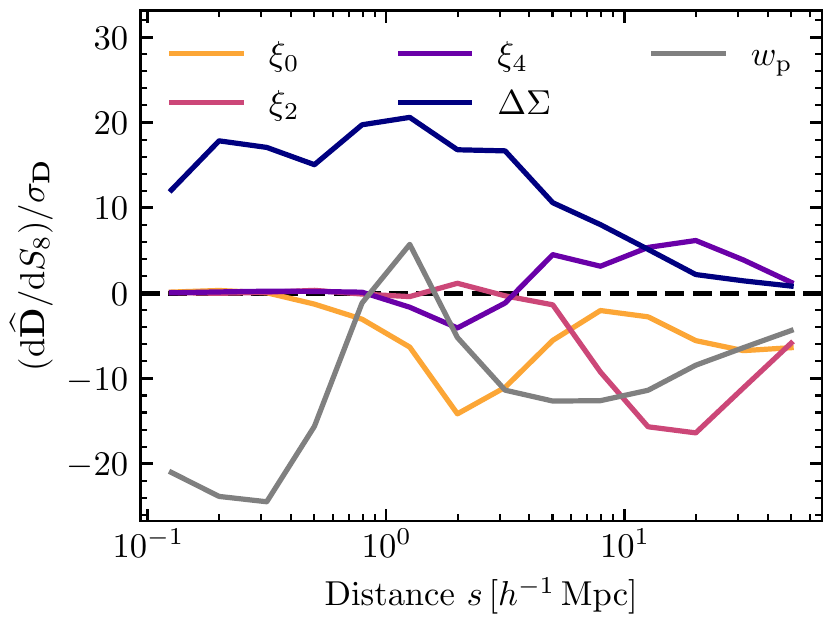}
    \caption{Estimates of the derivatives of the best-fit predictions as a function of $S_8$. The derivatives are derived from the mock data RSD fits presented in \protect\cite{Lange2022_MNRAS_509_1779}.}
    \label{fig:S_8_derivative}
\end{figure}

We now want to apply the above mentioned method to our application by masking the value of $S_8$. However, in simulation-based modelling, we can only make predictions for a handful of cosmologies and the predictions are inherently noisy due to finite simulation sizes. To overcome these problems, we slightly modify the method introduced above while keeping the main idea. In the mock analysis in \cite{Lange2022_MNRAS_509_1779}, we fit our model to a mock RSD vector and get best-fit model predictions $\widehat{\mathbf{D}}$ for all $40$ simulations. We then use the $40$ predictions to fit for a linear relation between $\widehat{\mathbf{D}}$ and the input $S_8$ value to estimate ${\rm d} \widehat{\mathbf{D}} / {\rm d} S_8$.

\begin{figure}
    \includegraphics{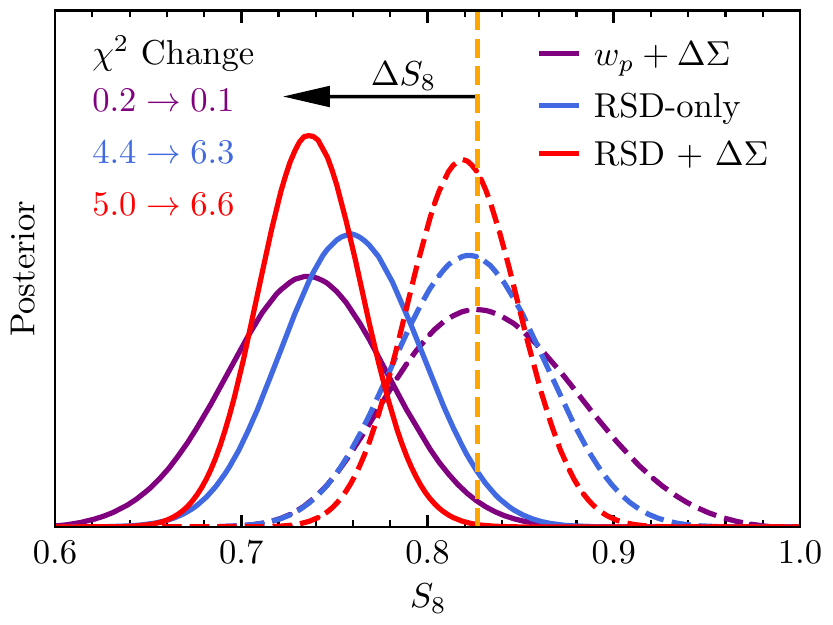}
    \caption{Change in the $S_8$ posterior constraints from the mock data set when applying the masking procedure outlined in section~\ref{sec:masking}. We show the original input $S_8$ value (yellow dashed vertical line) and the expected $S_8$ shift $\Delta S_8 = -0.1$ (black arrow). As expected, the posterior constraints obtained from the masked data (solid lines) are shifted by roughly $-0.1$ compared to the same constraints obtained from the unmasked data (dashed lines). In the upper right corner, we also indicate the change in the goodness of fit due to applying the masking.}
    \label{fig:mock_masking}
\end{figure}

The resulting derivatives are shown in Fig.~\ref{fig:S_8_derivative} and discussed further in section \ref{subsec:full_scale}. As expected, the figure indicates that $S_8$ and $\Delta\Sigma$ predictions are positively correlated. To mask $S_8$, we add the following offset to the data vector $\mathbf{D}$:
\begin{equation}
    \Delta \mathbf{D} (\Delta S_8) = \frac{{\rm d} \widehat{\mathbf{D}}}{{\rm d} S_8} \times \Delta S_8 \, .
\end{equation}

We test the above masking procedure on the mocks in the previous section. We choose a target shift of $\Delta S_8 = -0.1$. In Fig.~\ref{fig:mock_masking}, we show the shift in the $S_8$ posterior induced by the this choice. As expected, irrespective of whether we analyse $w_p$ and $\Delta\Sigma$, the redshift-space correlation function or the combination of lensing and redshift-space clustering, the $S_8$ posterior shifts by roughly $-0.1$ in $S_8$. Furthermore, as shown in the same figure, the goodness of fit is close to unaffected by the masking procedure. These findings on mock catalogues indicate that the masking procedure is expected to give a good performance in practical applications such as the one in the present work.

We then proceed to apply the masking procedure to the data. For each of the three redshift bins, we choose a random $\Delta S_8$ that is drawn from a uniform distribution in the range $[-0.075, +0.075]$. In principle, larger ranges would be ideal to erase any meaningful correlation between the masked and unmasked result. However, for our simulation-based analysis, we need to ensure that the value of $S_8$ that the masked data prefers is covered by the simulations. Thus, we cannot make the range for $\Delta S_8$ arbitrarily large. The range chosen here was selected as a compromise. Note that we apply the same $S_8$ masking shift for all analyses within each redshift bin. Therefore, even with the masking, we were able to judge the consistency between constraints coming from RSDs and lensing as well as between results from the NGC and the SGC data. The entire analysis in the next section was first performed and checked with the masked data and only unmasked after all authors agreed on all analysis choices.

\section{Results}
\label{sec:results}

We now proceed to apply the modelling framework to the observational galaxy clustering and galaxy--galaxy lensing data. Here, we concentrate on posterior constraints on cosmology. Constraints on the galaxy--halo connection are presented and discussed in appendix \ref{sec:galaxy-halo-connection}.

\subsection{Model fits}

\begin{figure*}
    \includegraphics{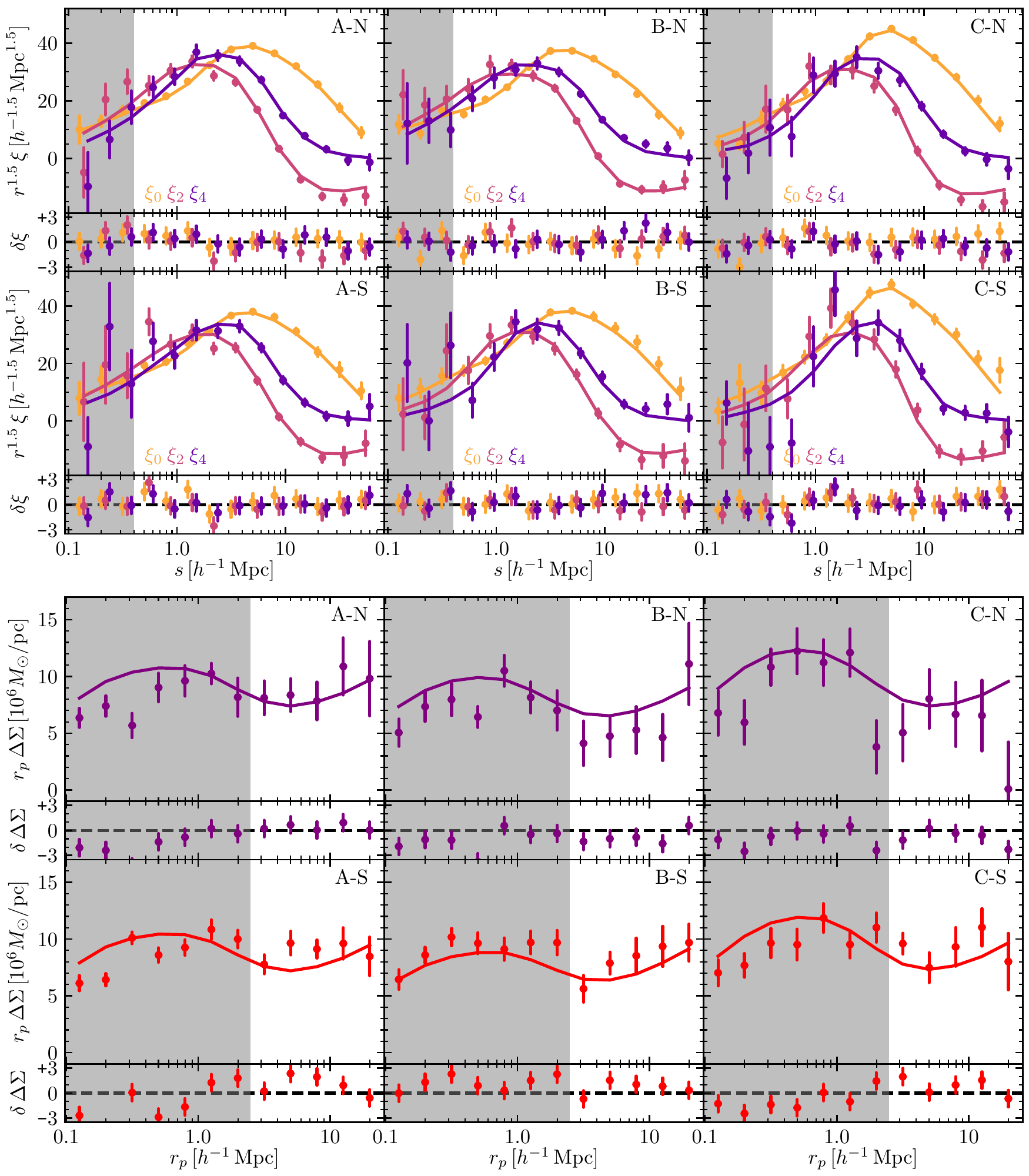}
    \caption{Data measurements and best-fit model predictions when analysing galaxy clustering and galaxy--galaxy lensing jointly. Upper panels show the clustering measurements and fits whereas the lower panels display the lensing measurements and models. Each panel indicates the sample displayed, e.g. ``A-S'' denotes sample A in the SGC area. For each set, the upper panels show the absolute measurements (data points) and predictions (solid lines) and the lower panels show the difference between the best-fit model predictions and the data in units of the observational uncertainty. As in Fig.~\ref{fig:mock_fit}, grey backgrounds indicate the ranges of the data that are not included in the fit. All fits use the {\sc Aemulus} simulation box {\sc B04} which provides the best fit to the combination of all observations.}
    \label{fig:data_fits}
\end{figure*}

\begin{figure}
    \includegraphics[width=\columnwidth]{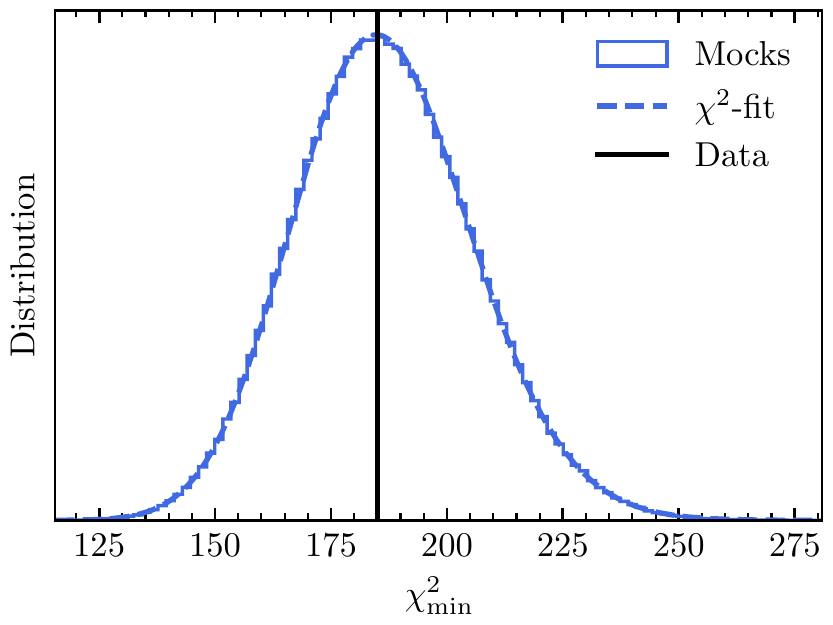}
    \caption{The distribution of best-fit $\chi^2$-values in perturbed mocks when fitting all data (solid blue), an analytic $\chi^2$-distribution with the same mean as the mocks (dashed blue) and the best-fit $\chi^2$ from the actual data (black). For each of the perturbed mocks, we fully marginalised over all $11$ galaxy--halo connection parameters.}
    \label{fig:chi_sq_dist_d}
\end{figure}

In Fig.~\ref{fig:data_fits}, we show the best-fit model predictions for each of the six samples. Each model was fitted to the redshift-space clustering and galaxy--galaxy lensing data jointly. The fits were marginalised over both the galaxy--halo connection parameters as well as over the $40$ {\sc Aemulus} simulations, i.e. cosmology. The minimum $\chi^2$ values are $23.0$, $39.2$ and $26.2$ ($32.5$, $18.5$ and $31.0$) for the three bins from low to high redshift in the NGC (SGC) with $39$ data points. When fitting all data with a single simulation, as shown in Fig.~\ref{fig:data_fits}, the best-fit $\chi^2$ is $185$. Although our HOD model has $11$ free parameters, the number of effective degrees of freedom of the galaxy--halo connection is smaller. We numerically determine this number by taking model predictions, randomly perturbing them according to the observational uncertainty and minimising $\chi^2$ over HOD parameter space. We find that the number of effective degrees of freedom of the HOD model with respect to fitting redshift-space clustering and lensing jointly is $\sim 8.5$. Similarly, while we have seven cosmological parameters that are varied in the {\sc Aemulus} simulations, the likelihood seems to be only a function of around two. Thus, we estimate to have around $\sim 2$ effective degrees of freedom in cosmology. Overall, when fitting redshift-space clustering and lensing, the number of effective degrees of freedom $N_{\rm dof}$ is approximately $39.0 - 8.5 - 2.0 = 28.5$ when fitting a single sample of galaxies and $6 \times (39.0 - 8.5) - 2.0 = 181$ when fitting all six galaxy samples. In Fig.~\ref{fig:chi_sq_dist_d}, we show the distribution of best-fit $\chi^2$ values for the mocks when fitting all six samples to redshift-space clustering and galaxy--galaxy lensing together with the best-fit $\chi^2$ from the data. Overall, we find $\chi^2_\nu = \chi^2 / N_{\rm dof} \approx 1$ for all samples. The largest $\chi^2_\nu$ value, $39.2 / 28.5$, has a $p$-value of $0.09$. Thus, overall, our model provides a good fit to all the available data.

\subsection{Cosmology}

\begin{figure*}
    \includegraphics{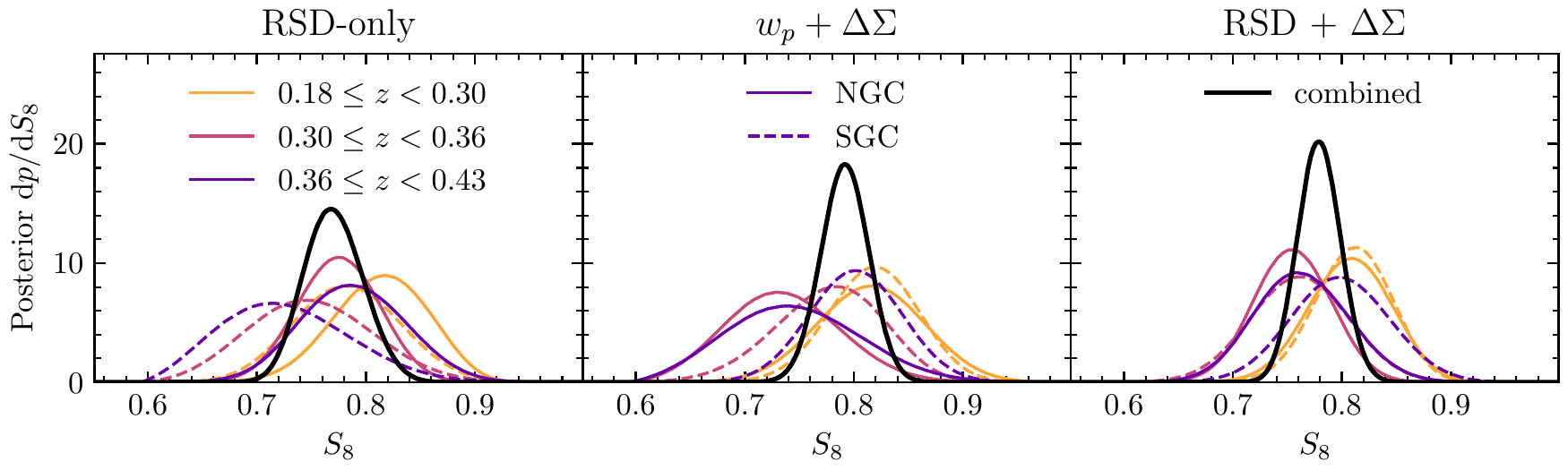}
    \caption{Posterior constraints on $S_8$ from the combination of $w_p$ and $\Delta\Sigma$ (left), redshift-space distortions (middle) and the combination of redshift-space clustering and lensing (right). In each panel, we show constraints from different redshift bins (different coloured lines) from the NGC (solid) and SGC (dashed) area. Additionally, we show constraints when combining all six LOWZ data samples (black solid).}
    \label{fig:s_8}
\end{figure*}

\begin{table}
    \centering
    \begin{tabular}{cccc}
        sample & $w_p + \Delta\Sigma$ & RSD-only & RSD + $\Delta\Sigma$ \\
        \hline\hline
        A NGC & $0.815 \pm 0.049$ & $0.813 \pm 0.043$ & $0.807 \pm 0.038$ \\
        A SGC & $0.818 \pm 0.041$ & $0.783 \pm 0.048$ & $0.810 \pm 0.035$ \\
        B NGC & $0.735 \pm 0.052$ & $0.774 \pm 0.037$ & $0.754 \pm 0.035$ \\
        B SGC & $0.776 \pm 0.049$ & $0.750 \pm 0.056$ & $0.761 \pm 0.045$ \\
        C NGC & $0.746 \pm 0.060$ & $0.789 \pm 0.048$ & $0.763 \pm 0.043$ \\
        C SGC & $0.802 \pm 0.042$ & $0.727 \pm 0.058$ & $0.798 \pm 0.045$ \\
        \hline
        combined & $0.792 \pm 0.022$ & $0.771 \pm 0.027$ & $0.779 \pm 0.020$ \\
    \end{tabular}
    \caption{Posterior constraints on the cosmological parameter $S_8$ as a function of the galaxy sample analysed (different rows) and the observational constraints (different columns).}
    \label{tab:s_8}
\end{table}

\begin{figure}
    \includegraphics{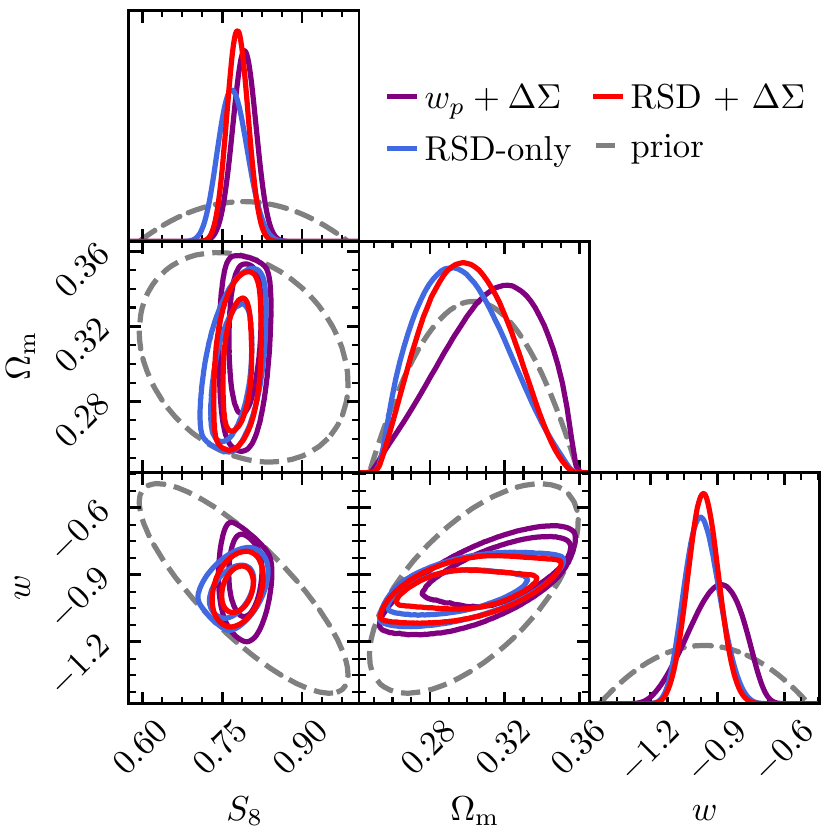}
    \caption{Cosmological constraints from BOSS LOWZ when analysing all six samples in this work jointly. We show constraints derived from a combination of projected galaxy clustering and galaxy--galaxy lensing (purple), redshift-space clustering (blue) and the combination of redshift-space clustering and lensing (red). We also show the prior imposed by the {\sc Aemulus} simulations (grey).}
    \label{fig:posterior_d_l}
\end{figure}

Fig.~\ref{fig:s_8} shows our constraints on the cosmological parameter $S_8$. Results are presented  for all six samples individually. We also distinguish between  redshift-space clustering-only fits, the combination of projected clustering and galaxy--galaxy lensing, and joint redshift-space clustering and lensing fits. Overall, the different samples show good agreement for the value of $S_8$. Furthermore, the RSD-only fits and the joint projected clustering plus lensing fits are also in good agreement. When combining all six galaxy samples, we obtain $S_8 = 0.792 \pm 0.022$ from gravitational lensing and $S_8 = 0.771 \pm 0.027$ from redshift-space clustering. We also investigate how the redshift-space clustering result depends on the scales considered. By default, we consider all scales larger than $400 \, \hkpc$. When only analysing scales larger than $6.3 \, \hmpc$, we obtain $S_8 = 0.806 \pm 0.042$. Finally, when looking at the combination of reshift-space clustering and lensing for all six samples, we obtain $S_8 = 0.779 \pm 0.020$. The derived constraints on $S_8$ are also listed in Table~\ref{tab:s_8}, for convenience. In Fig.~\ref{fig:posterior_d_l}, we show the full cosmology constraints on $S_8$, $\Omega_{\rm m}$ and $w$ when considering all six samples. In addition to constraints on $S_8$, we infer $w = -0.915 \pm 0.113$, $-0.967 \pm 0.076$ and $-0.963 +/- 0.069$ for observations of $w_p + \Delta\Sigma$, RSD-only and RSD $+ \Delta\Sigma$, respectively. On the other hand, we do not obtain strong constraints on $\Omega_{\rm m}$ and are instead dominated by the {\sc Aemulus} prior.

\subsection{Lensing amplitudes derived from different lensing data sets}
\label{subsec:lensing_amplitudes}

\begin{figure*}
    \includegraphics{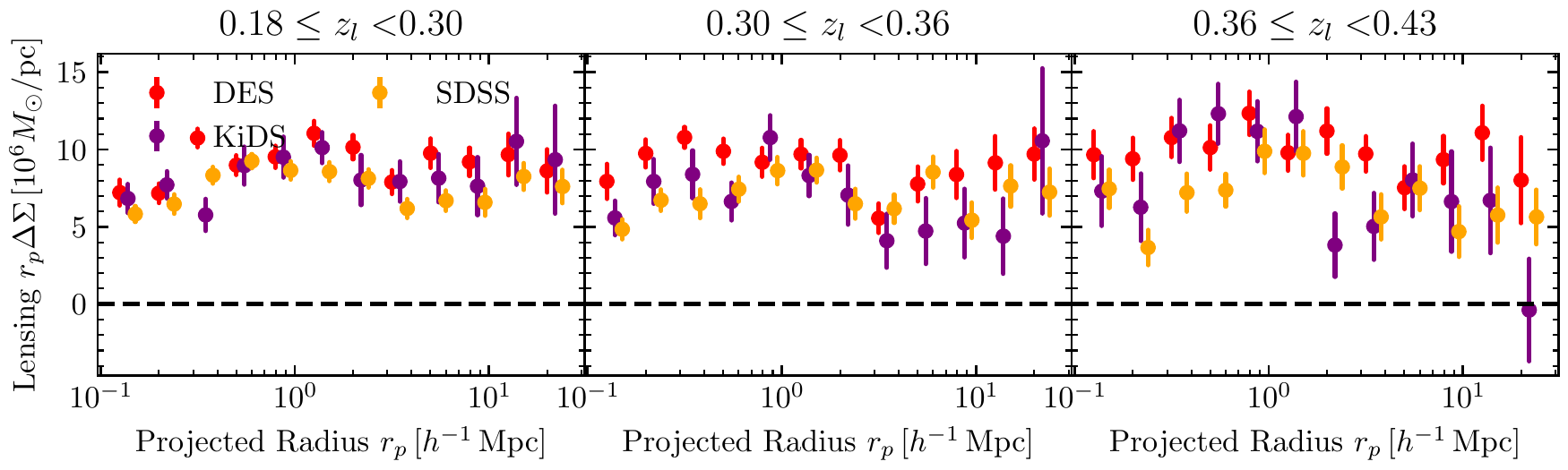}
    \caption{Galaxy--galaxy lensing measurements from cross-correlating BOSS LOWZ targets with lensing catalogues from DES (red), KiDS (purple), and SDSS (orange). Different panels correspond to the three different redshift-binned samples analysed in this work. Measurements are slightly offset in the $x$-direction for clarity.}
    \label{fig:lwb}
\end{figure*}

Here, we compare the measured galaxy--galaxy lensing amplitudes between KiDS and DES. Additionally, we contrast them with galaxy--galaxy lensing measurements obtained with SDSS lensing catalogues \citep{Singh2019_MNRAS_482_785}. In this section, we use lensing measurements that extend to smaller radial scales than used in the fiducial cosmology analysis. The lensing amplitude $\Delta\Sigma$ is a physical quantity that should only depend on lens properties and be independent of the shape catalogue used. In a recent study, \cite{Leauthaud2022_MNRAS_510_6150} used this insight to compare the results of different lensing catalogues, including SDSS, KiDS and DES, finding good agreement between all of them within the statistical and systematic uncertainties. However, the findings of \cite{Leauthaud2022_MNRAS_510_6150} are based on older DES Y1 and KiDS-450 data whereas our new DES and KiDS lensing measurements have substantially more statistical constraining power and reduced systematics. We compare the SDSS, DES and KiDS lensing measurements in Fig.~\ref{fig:lwb}. The lensing measurements used for our cosmology analysis are limited to scales where boost factors corrections due to physical lens--source associations are unimportant. Here, we extend the measurements to smaller radial scales and, thus, include boost factor corrections. Overall, we see that the DES Y3 lensing measurements tend to be higher than the SDSS and KiDS measurements. To estimate the statistical significance of the difference in the lensing amplitudes, we follow the approach of \cite{Leauthaud2022_MNRAS_510_6150}. In particular, we first determine an overall lensing normalisation $A$ by fitting the observed lensing amplitude $\Delta\Sigma_{\rm obs}$ with a template $\Delta\Sigma_{\rm template}$. Afterwards, we compare the normalisations between the different lensing amplitudes. For the template, we choose the best-fit lensing prediction when analyzing the combination of galaxy redshift-space clustering and galaxy--galaxy lensing with the simulation box {\sc B04}, though the exact choice of template does not strongly affect the results.

\begin{figure}
    \includegraphics{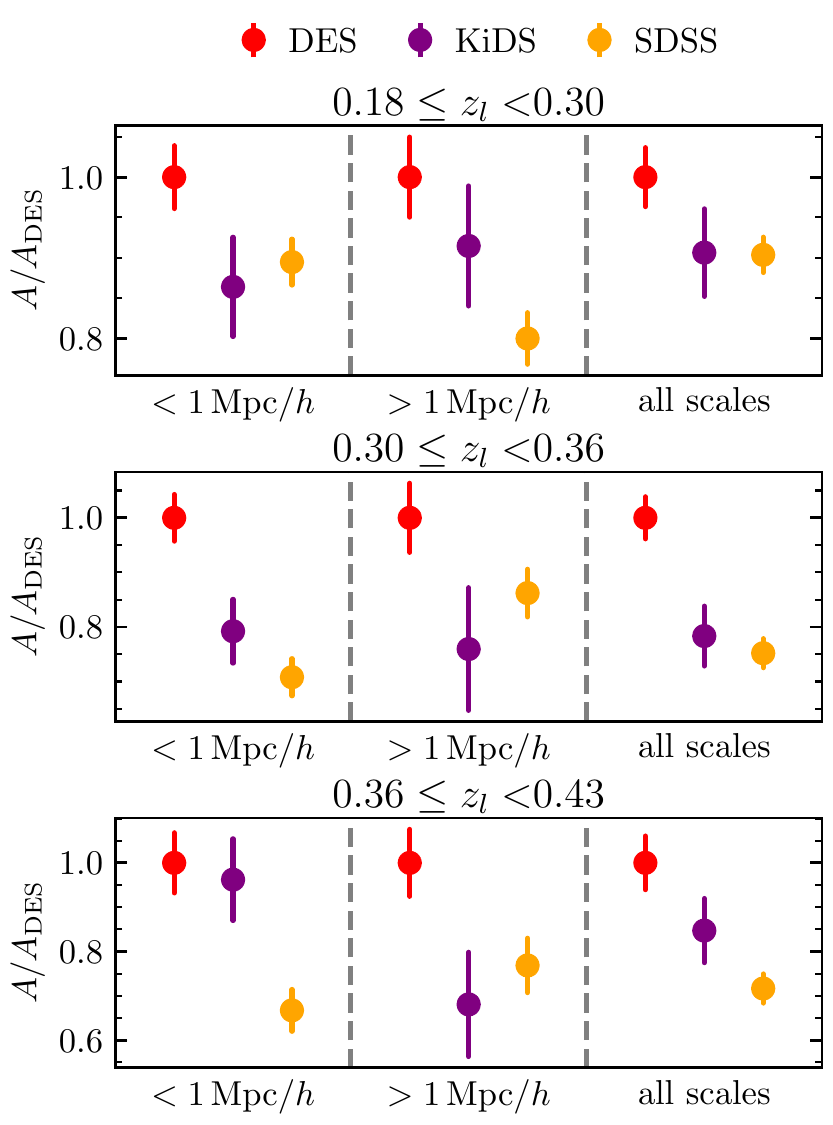}
    \caption{Lensing amplitudes averaged over different scales as a function of the lensing data set, the lens galaxy sample and the scale range. Three different redshift bins are shown from low redshift (top) to higher redshift (bottom).}
    \label{fig:lwb_amplitudes}
\end{figure}

The results are shown in Fig~\ref{fig:lwb_amplitudes}. When comparing DES and SDSS on scales $r_p > 1 \hmpc$, we find that SDSS lensing amplitudes are $(20 \pm 6) \%$, $(14 \pm 8) \%$ and $(23 \pm 10) \%$ lower on average than the DES amplitudes for samples A, B and C, respectively. When doing the same comparison for KiDS, we find $(-13 \pm 9) \%$, $(14 \pm 16) \%$ and $(13 \pm 19) \%$, i.e. KiDS and SDSS agree well on large scales. We note that the quoted uncertainties are the statistical uncertainties only and do not account for systematic uncertainties. The latter should be dominated by the SDSS photometric redshift calibration and is of order $6\%$ \citep{Singh2019_MNRAS_482_785}. Taking into account this systematic uncertainty, none of the offsets at large scales are statistically significant. On the other hand, differences on smaller scales have a stronger significance but are subject to uncertainties regarding boost factor estimates \citep{Leauthaud2017_MNRAS_467_3024}. We leave an in-depth comparison of lensing amplitudes measured with different lensing data sets to a future study with new lenses from the DESI survey.

\section{Discussion}
\label{sec:discussion}

To our knowledge, the current work represents the first full-scale combined analysis of redshift-space clustering and galaxy--galaxy lensing. Using this approach and combining all LOWZ samples, we obtain a $\sim 2.5 \%$ constraint on $S_8$,  one of the most stringent constraints on $S_8$ to date \citep{Abdalla2022_JHEAp_34_49}.

\subsection{\texorpdfstring{$S_8$-tension}{S8-tension}}

\begin{figure}
    \includegraphics{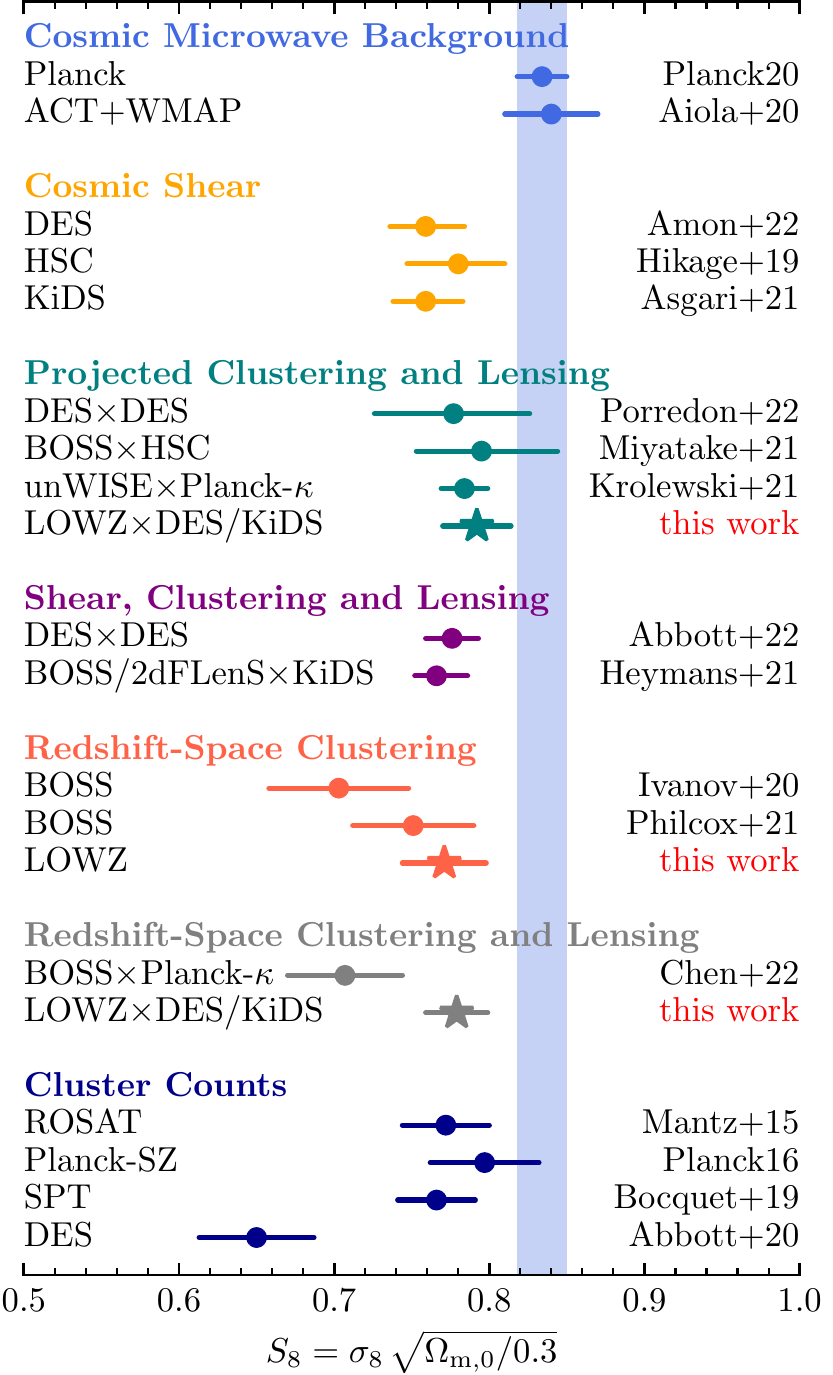}
    \caption{Comparison of different literature constraints on $S_8$ against the results derived in this work \protect\citep[results from][]{Mantz2015_MNRAS_446_2205, PlanckCollaboration2016_AA_594_24, Hikage2019_PASJ_71_43, Bocquet2019_ApJ_878_55, PlanckCollaboration2020_AA_641_6, Aiola2020_JCAP_12_047, Ivanov2020_JCAP_05_042, Abbott2020_PhRvD_102_3509, Asgari2021_AA_645_104, Krolewski2021_JCAP_12_028, Porredon2022_PhRvD_106_3530, Miyatake2022_PhRvD_106_3520, Heymans2021_AA_646_140, Amon2022_PhRvD_105_3514, Philcox2022_PhRvD_105_3517, Abbott2022_PhRvD_105_3520, Chen2022_JCAP_07_041}. Blue band indicates the prediction of Planck2020.}
    \label{fig:s8_comparison}
\end{figure}

In this work, we present two roughly independent constraints on the growth of structure amplitude $S_8$ based on the analysis of redshift-space clustering and the combination of projected clustering and galaxy--galaxy lensing, $S_8 = 0.771 \pm 0.027$ and $0.792 \pm 0.022$, respectively. By contrast, the most recent Planck2020 CMB analysis prefers $S_8 = 0.834 \pm 0.016$. This represents a $\sim 2 \sigma$ discrepancy in both cases. While this offset is not statistically significant, our results follow a long-standing trend whereby low-redshift probes of cosmic structure growth infer a lower value for $S_8$ than studies of the CMB. In Fig.~\ref{fig:s8_comparison}, we compare our results against CMB results \citep{PlanckCollaboration2020_AA_641_6, Aiola2020_JCAP_12_047} and other low-redshift large-scale structure studies, including cosmic shear \citep{Hikage2019_PASJ_71_43, Asgari2021_AA_645_104, Amon2022_PhRvD_105_3514}, the combination of projected galaxy clustering and lensing cross-correlation \citep{Krolewski2021_JCAP_12_028, Porredon2022_PhRvD_106_3530, Miyatake2022_PhRvD_106_3520}, so-called $3\times2$pt studies \citep{Heymans2021_AA_646_140, Abbott2022_PhRvD_105_3520}, redshift-space clustering \citep{Ivanov2020_JCAP_05_042, Philcox2022_PhRvD_105_3517}, the combination of redshift-space clustering and lensing cross-correlation \citep{Chen2022_JCAP_07_041} as well as cluster counts \citep{Mantz2015_MNRAS_446_2205, PlanckCollaboration2016_AA_594_24, Bocquet2019_ApJ_878_55, Abbott2020_PhRvD_102_3509}. Overall, our results are in excellent agreement with other low-redshift probes which also prefer $S_8 \sim 0.76$. Our analysis is the first to analyse BOSS LOWZ galaxies cross-correlated with KiDS-1000 and DES Y3 to fit for cosmology. Additionally, our constraints from redshift-space clustering are primarily derived from scales not analysed in conventional large scale-only RSD studies  \citep[e.g.,][]{Ivanov2020_JCAP_05_042, Philcox2022_PhRvD_105_3517}, which we have achieved through a simulation-based model that marginalises over uncertainties in galaxy assembly bias and other astrophysical effects. We note that two other recent full-scale RSD studies of the BOSS CMASS \citep{Zhai2022_arXiv_2203_8999} and the eBOSS LRG samples \citep{Chapman2022_MNRAS_516_617} also find a lower growth of structure amplitude than predicted by Planck2020. Overall, our results add further evidence for the existence of an $S_8$-tension between low redshift and CMB data and to the evidence that this discrepancy is not limited to studies involving gravitational lensing.

\subsection{Lensing is low}
\label{subsec:lensing_is_low}

For a given set of cosmological parameters, the observed galaxy clustering amplitude makes precise predictions for the galaxy--galaxy lensing amplitude, even after marginalising over uncertainties in the galaxy--halo connection \citep[e.g.][]{Yoo2006_ApJ_652_26, Cacciato2009_MNRAS_394_929, Leauthaud2017_MNRAS_467_3024}. Recently, several studies have shown that the lensing amplitude is significantly over-predicted when assuming the best-fit Planck2020 cosmological parameters. This discrepancy is most significant on small scales, $\rp < 5 \, \hmpc$ where the difference is around $30\%$ \citep{Leauthaud2017_MNRAS_467_3024, Lange2019_MNRAS_488_5771, Yuan2020_MNRAS_493_5551, Lange2021_MNRAS_502_2074, Amon2023_MNRAS_518_477}. However, the matter distribution on such small scales is also affected by baryonic feedback effects which are often not explicitly modelled. This may contribute to the apparent lensing-is-low effect on small scales \citep{Leauthaud2017_MNRAS_467_3024, Lange2019_MNRAS_488_5771, Amodeo2021_PhRvD_103_3514}, which we do not analyse here.

On large scales, the lensing measurements have increased uncertainties and therefore it is less clear to what extent a similar lensing-is-low tension exists at those scales, as well. Recently, \cite{Lange2021_MNRAS_502_2074}, using BOSS LOWZ clustering and SDSS galaxy--galaxy lensing measurements, also find a statistically significant difference of $\sim 30 - 35 \%$ on large scales. Recall that the lensing predictions at fixed projected clustering roughly scale as $S_8$ \citep{Yoo2006_ApJ_652_26}. Thus, studies cross-correlating BOSS LOWZ galaxies with SDSS shape catalogues and inferring $S_8 \sim 0.71 \pm 0.03$ \citep{Wibking2020_MNRAS_492_2872} and $\sim 0.71 \pm 0.04$ \citep{Singh2020_MNRAS_491_51}, considerably below the Planck CMB prediction, also corroborate the existence of a lensing-is-low problem. More recently, \cite{Amon2023_MNRAS_518_477} re-evaluated the lensing-is-low tension using updated DES and KiDS lensing measurements, finding a difference at the level of $15\%$ for LOWZ on large scales. Overall, their results are offset with respect to with the Planck CMB prediction at the $\sim 2\sigma$  level. Finally, our results also imply a mild, $2\sigma$ tension with the Planck2020 results but not at the level reported in earlier studies relying on SDSS lensing measurements \citep{Singh2020_MNRAS_491_51, Wibking2020_MNRAS_492_2872, Lange2021_MNRAS_502_2074}.

As shown in section~\ref{subsec:lensing_amplitudes}, we find that for LOWZ galaxies the SDSS lensing catalogues imply lower lensing amplitudes for the same lens samples than the DES Y3 catalogues. We leave a detailed investigation of the statistical significance of this finding to future work. Nonetheless, this difference helps explain why earlier studies based on SDSS lensing measurements \citep{Wibking2020_MNRAS_492_2872, Singh2020_MNRAS_491_51, Lange2021_MNRAS_502_2074} find stronger levels of tension with Planck CMB predictions than \cite{Amon2023_MNRAS_518_477} and the present study.

\subsection{Full-scale studies}
\label{subsec:full_scale}

Using a full-scale approach, growth of structure constraints from redshift-space distortions become competitive with leading gravitational lensing studies. Recent large scale-only, full-shape studies using BOSS LOWZ and CMASS achieve a $\sim 0.045$ uncertainty on $S_8$ \citep{Ivanov2020_JCAP_05_042, Philcox2022_PhRvD_105_3517}, whereas here we obtain a $0.027$ constraint. Furthermore, in this work we only use the BOSS LOWZ sample and do not analyse the two times larger BOSS CMASS sample, unlike the aforementioned large scale-only studies. In Fig.~\ref{fig:S_8_derivative}, we show the derivative of the RSD multipoles with respect to changes in $S_8$ after fully marginalising over galaxy--halo connection parameters. Thus, this figure indicates where cosmological constraints from full-scale studies are coming from. It indeed suggests that significant information on $S_8$ derives from scales smaller than $10 \, \hmpc$. At the same time, the figure also confirms the finding in \cite{Lange2022_MNRAS_509_1779} that little to no cosmological information is contained for RSD multipoles below $s \sim 1 \, \hmpc$. We also repeat the RSD analysis using only scales above $6.3 \, \hmpc$, finding that our constraints on $S_8$ degrade by $60\%$. Additionally, we find that excluding the smallest scales from our analysis leads to a constraint on $S_8$ that, while being higher \citep[also see][]{Lange2022_MNRAS_509_1779, Chapman2022_MNRAS_516_617, Zhai2022_arXiv_2203_8999}, is in good agreement with the result from the default analysis. Overall, our results suggest that full-scale studies have the potential of improving cosmological constraints from redshift-space clustering by a factor of $2 - 3$ over traditional large scale-only full-shape studies, even after marginalising over complex galaxy--halo connection models \citep[also see][]{Lange2022_MNRAS_509_1779}. In addition to placing independent, high-precision constraints on the $S_8$-tension, analyses of the non-linear regime could also enable precise tests of modifications to gravity in the future \citep{Blake2020_AA_642_158, Alam2021_JCAP_11_050}.

Regarding the combination of projected clustering and galaxy--galaxy lensing, we model the lensing signal down to $2.5 \, \hmpc$. Nominally, this is not a significant improvement to, for example, the recent analysis by \cite{Porredon2022_PhRvD_106_3530}, where scales down to $6 \, \hmpc$ are considered. However, the analysis of \cite{Porredon2022_PhRvD_106_3530} marginalises over a free point-mass term, which scales as $\Delta\Sigma \propto \rp^{-2}$ and is designed to remove systematics in the modelling of non-linear scales. As described in \cite{Prat2022_PhRvD_105_3528}, this reduces the signal-to-noise ratio of the lensing measurements analysed in \cite{Porredon2022_PhRvD_106_3530} from $67$ to $32$ and primarily affects small scales. Thus, \cite{Porredon2022_PhRvD_106_3530} and similar analyses marginalising over a point-mass term, e.g. \cite{Singh2020_MNRAS_491_51} and \cite{Wibking2020_MNRAS_492_2872}, are by design less sensitive to the predicted lensing amplitude on small scales relative to studies without point-mass marginalisation. In this work, we do not marginalise over a point-mass term since we argue that our simulation-based modelling framework and complex galaxy--halo connection model should already accurately model scales down to $2.5 \, \hmpc$, thereby allowing us to leverage this additional information without the sacrifice in signal-to-noise that results from point-mass marginalisation.

Currently we do not explore even smaller scales in order to avoid the need to model baryonic feedback processes \citep{Lange2019_MNRAS_488_5771, Amodeo2021_PhRvD_103_3514}. However, the lensing measurements on scales down to $0.1 \, \hmpc$ have a signal-to-noise ratio of $57$ compared to $26$ when limiting the analysis to scales $> 2.5 \, \hmpc$. Thus, if we would be able to empirically constrain and marginalise over baryonic feedback processes, we could potentially get even more stringent growth-of-structure constraints. Unfortunately, galaxy clustering alone is unlikely to place strong constraints on baryonic feedback, thus marginalising over flexible baryonic feedback models with the present data is unlikely to result in more stringent $S_8$ constraints. However, combining galaxy clustering and lensing with data on the baryon distribution around galaxies, such as measurements of the Sunyaev--Zel'dovich \citep{Schaan2021_PhRvD_103_3513, Amodeo2021_PhRvD_103_3514}, may break that degeneracy, and could unlock the constraining power of small-scale lensing measurements for cosmology.

The constraining power of full-scale studies might be further improved by using larger simulations with reduced sample variance. The current {\sc Aemulus} simulations have a volume of $(\sim 1 \, \hgpc^3)$ each, roughly the same as the total LOWZ volume analysed here. Since our analysis approach incorporates uncertainties in simulation predictions, constraints would likely become more stringent with larger simulations such as the AbacusSummit simulation suite \citep{Maksimova2021_MNRAS_508_4017}. On the observational side, most current spectroscopic large-scale structure surveys are designed to be cosmic variance limited on large scales, i.e. up to $k \sim 0.2 \, h \, \mathrm{Mpc}^{-1}$. Due to this design choice, highly non-linear scales are currently mostly dominated by shot noise. Future galaxy surveys with a higher sampling density might improve the constraining power of non-linear scales further for the same observational volume \citep{Dawson2022_arXiv_2203_7291}. Such high-density surveys would also be ideal targets for multi-tracer studies, another method to reduce the impact of cosmic variance \citep{McDonald2009_JCAP_10_007}.

At the same time, simulations and modelling frameworks need to be developed further. For example, current gravity solvers predict an up to $1 \%$ different halo (matter) clustering amplitude at $k \sim 1 (10) \, h \, \mathrm{Mpc}^{-1}$ respectively at low redshifts \citep{Grove2022_MNRAS_515_1854}. With increasing precision of large-scale structure measurements, these differences might soon dominate uncertainties in the cosmological interpretation of non-linear scales. Furthermore, more work needs to be done to ensure that the galaxy--halo connection models used to analyse the data are sufficiently general and do not bias cosmology results. While several cosmology recovery tests \citep{Lange2022_MNRAS_509_1779, Zhai2022_arXiv_2203_8999} have been performed on SHAM mock galaxy catalogues, including in this work, tests on additional, more complex galaxy models, including semi-analytic models and hydrodynamic simulations \citep[see][for a review]{Wechsler2018_ARAA_56_435}, are needed. At the same time, marginalising over more physically-motivated galaxy--halo connection models than HODs such as the {\sc UniverseMachine} \citep{Behroozi2019_MNRAS_488_3143} or {\sc Emerge} \citep{Moster2018_MNRAS_477_1822} might reduce cosmological uncertainties. We leave such investigations to future work.

\section{Conclusion}
\label{sec:conclusions}

In this work, we present a novel simulation-based joint cosmological analysis of galaxy redshift-space clustering and galaxy--galaxy lensing. Our work extends previous simulation-based full-scale studies \citep[e.g.][]{Wibking2020_MNRAS_492_2872, Miyatake2022_PhRvD_106_3520, Lange2022_MNRAS_509_1779, Zhai2022_arXiv_2203_8999} in several directions. For example, this is the first full-scale cosmological study involving gravitational lensing that also models galaxy assembly bias. Furthermore, this is the first simulation-based cosmological full-scale study that uses the most recent DES Y3 and KiDS-1000 gravitational lensing measurements. Most importantly, we also present the first joint full-scale analysis of redshift-space clustering and galaxy--galaxy lensing.

Our analysis incorporates a complex galaxy--halo connection model, including the effects of galaxy assembly bias as well as central and satellite velocity bias. Our best-fit model is able to provide a good fit to the observed galaxy clustering and lensing amplitudes over a wide range of scales, down to $2.5 \, \hmpc$ for lensing and $0.4 \, \hmpc$ for clustering. Our main result are new, highly competitive constraints on the cosmic growth of structure, particularly $S_8$. These findings can be summarised as follows.

\begin{itemize}
    \item When analysing the combination of projected clustering and galaxy--galaxy lensing, we infer $S_8 = 0.792 \pm 0.022$ while for redshift-space clusetering we infer $S_8 = 0.771 \pm 0.027$. Finally, combining redshift-space clustering and galaxy--galaxy lensing, we find $S_8 = 0.779 \pm 0.020$.
    \item We find good agreement regarding $S_8$ between multiple independent galaxy samples. Similarly, constraints derived only from redshift-space clustering are consistent with those relying on gravitational lensing.
    \item When repeating our analysis of redshift-space clustering using only larger scales above $s > 6.3 \, \hmpc$ instead of $400 \, \hkpc$, we achieve statistically consistent results, but our constraining power on $S_8$ degrades by $60\%$.
    \item Our results favour a value for $S_8$ below the best-fit value inferred by the Planck2020 CMB analysis, $S_8 = 0.834$. This is in agreement with results from other analyses of the low-redshift Universe, the so-called $S_8$-tension \citep[see][for a review]{Abdalla2022_JHEAp_34_49}.
    \item Similar to other recent full-scale studies of galaxy redshift-space clustering \citep{Chapman2022_MNRAS_516_617, Zhai2022_arXiv_2203_8999, Yuan2022_MNRAS_515_871}, we find evidence for an $S_8$-tension with predictions from Planck2020, even without incorporating gravitational lensing data.
\end{itemize}

Our analysis also highlights the strong constraining power of full-scale studies over similar analyses targeting only large scales. Particularly, our constraints on $S_8$ using only redshift-space clustering are a factor of two more stringent than recent large scale-only studies of BOSS galaxy redshift-space clustering \citep{Ivanov2020_JCAP_05_042, Philcox2022_PhRvD_105_3517}, despite only using a small fraction of the data compared to these other works. Similarly, our constraints derived from the combination of projected galaxy clustering and galaxy--galaxy lensing are significantly more stringent than, for example, a recent DES Y3 analysis using those observables \citep{Porredon2022_PhRvD_106_3530}. We attribute part of this improvement to considering scales down to $2.5 \, \hmpc$ for the lensing signal. Furthermore, we argue that our complex modelling framework alleviates the need to marginalise over a point mass, further increasing sensitivity to high-$S/N$ non-linear scales.

The current study represents a new advance in full-scale cosmological studies. We anticipate building upon this in the future in several respects. In the present study, we do not use scales below $2.5 \, \hmpc$ since our galaxy model model does not include baryonic feedback \citep{Leauthaud2017_MNRAS_467_3024, Lange2019_MNRAS_488_5771}. However, the signal-to-noise ratio of the lensing amplitude more than doubles when considering scales down to $\rp = 0.1 \, \hmpc$. Observations of the Sunyaev--Zel'dovich effect can place independent constraints on the strength of baryonic feedback \citep{Schaan2021_PhRvD_103_3513, Amodeo2021_PhRvD_103_3514}; thus, a combined analysis of clustering, lensing and the Sunyaev--Zel'dovich effect could improve constraining power even further. We aim to apply the full-scale approach to upcoming data from the DESI survey using the high-resolution, large-volume AbacusSummit simulations \citep{Maksimova2021_MNRAS_508_4017} instead of {\sc Aemulus} for the modelling. This is also expected to improve cosmological constraints. At the same time, in light of this increased constraining power, more tests on complex, highly-realistic galaxy mock catalogues are needed to verify the robustness of full-scale constraints. Another avenue for full-scale studies is to provide tight priors on galaxy bias models to be used with large-scale hybrid effective field theory cosmology studies \citep[see][for a recent example]{Kokron2022_MNRAS_514_2198}.

\section*{Acknowledgements}

We thank the {\sc Aemulus} collaboration and the {\sc UnitSims} team for making their simulations publicly available as well as Sandy Yuan, Jeremy Tinker, and Zhongxu Zhai for interesting discussions on several aspects of this analysis. We also thank Sukhdeep Singh for providing the SDSS lensing measurements discussed in section~\ref{subsec:lensing_is_low}.

We acknowledge use of the lux supercomputer at UC Santa Cruz, funded by NSF MRI grant AST 1828315. This work was partially supported by the U.S. Department of Energy, Office of Science, Office of High Energy Physics under Award Number DE-SC0019301. JUL received support from a fellowship from the Leinweber Center for Theoretical Physics and from a Stanford-Santa Cruz Fellowship including support from the Kavli Institute for Particle Astrophysics and Cosmology. AL acknowledges support from the David and Lucille Packard foundation, and from the Alfred P. Sloan foundation. HG acknowledges the support from the National Natural Science Foundation of China (Nos. 11833005, 11922305). Work done by APH was supported by the U.S. Department of Energy, Office of Science, Office of Nuclear Physics, under contract DE-AC02-06CH11357. FvdB is supported by the National Aeronautics and Space Administration through Grant No. 19-ATP19-0059 issued as part of the Astrophysics Theory Program.

This work made use of the following software packages: {\sc matplotlib} \citep{Hunter2007_CSE_9_90}, {\sc SciPy}, {\sc NumPy} \citep{vanderWalt2011_CSE_13_22}, {\sc Astropy} \citep{AstropyCollaboration2013_AA_558_33}, {\sc Colossus} \citep{Diemer2015_ascl_soft_1016}, {\sc halotools} \citep{Hearin2017_AJ_154_190}, {\sc MultiNest} \citep{Feroz2008_MNRAS_384_449, Feroz2009_MNRAS_398_1601, Feroz2019_OJAp_2_10}, {\sc PyMultiNest} \citep{Buchner2014_AA_564_125}, {\sc scikit-learn} \citep{Pedregosa2011_JMLR_12_2825}, {\sc emcee} \citep{ForemanMackey2013_PASP_125_306}, {\sc UltraNest} \citep{Buchner2021_JOSS_6_3001}, {\sc Spyder} and {\sc Setzer}.

This project used public archival data from the Dark Energy Survey (DES). Funding for the DES Projects has been provided by the U.S. Department of Energy, the U.S. National Science Foundation, the Ministry of Science and Education of Spain, the Science and Technology FacilitiesCouncil of the United Kingdom, the Higher Education Funding Council for England, the National Center for Supercomputing Applications at the University of Illinois at Urbana-Champaign, the Kavli Institute of Cosmological Physics at the University of Chicago, the Center for Cosmology and Astro-Particle Physics at the Ohio State University, the Mitchell Institute for Fundamental Physics and Astronomy at Texas A\&M University, Financiadora de Estudos e Projetos, Funda{\c c}{\~a}o Carlos Chagas Filho de Amparo {\`a} Pesquisa do Estado do Rio de Janeiro, Conselho Nacional de Desenvolvimento Cient{\'i}fico e Tecnol{\'o}gico and the Minist{\'e}rio da Ci{\^e}ncia, Tecnologia e Inova{\c c}{\~a}o, the Deutsche Forschungsgemeinschaft, and the Collaborating Institutions in the Dark Energy Survey.

The Collaborating Institutions are Argonne National Laboratory, the University of California at Santa Cruz, the University of Cambridge, Centro de Investigaciones Energ{\'e}ticas, Medioambientales y Tecnol{\'o}gicas-Madrid, the University of Chicago, University College London, the DES-Brazil Consortium, the University of Edinburgh, the Eidgen{\"o}ssische Technische Hochschule (ETH) Z{\"u}rich,  Fermi National Accelerator Laboratory, the University of Illinois at Urbana-Champaign, the Institut de Ci{\`e}ncies de l'Espai (IEEC/CSIC), the Institut de F{\'i}sica d'Altes Energies, Lawrence Berkeley National Laboratory, the Ludwig-Maximilians Universit{\"a}t M{\"u}nchen and the associated Excellence Cluster Universe, the University of Michigan, the National Optical Astronomy Observatory, the University of Nottingham, The Ohio State University, the OzDES Membership Consortium, the University of Pennsylvania, the University of Portsmouth, SLAC National Accelerator Laboratory, Stanford University, the University of Sussex, and Texas A\&M University.

Based in part on observations at Cerro Tololo Inter-American Observatory, National Optical Astronomy Observatory, which is operated by the Association of Universities for Research in Astronomy (AURA) under a cooperative agreement with the National Science Foundation.

Based on observations made with ESO Telescopes at the La Silla Paranal Observatory under programme IDs 177.A-3016, 177.A-3017, 177.A-3018 and 179.A-2004, and on data products produced by the KiDS consortium. The KiDS production team acknowledges support from: Deutsche Forschungsgemeinschaft, ERC, NOVA and NWO-M grants; Target; the University of Padova, and the University Federico II (Naples).

We use the gold sample of weak lensing and photometric redshift measurements from the fourth data release of the Kilo-Degree Survey \citep{Kuijken2019_AA_625_2, Wright2020_AA_637_100, Hildebrandt2021_AA_647_124, Giblin2021_AA_645_105} (Kuijken et al. 2019), hereafter referred to as KiDS-1000. Cosmological parameter constraints from KiDS-1000 have been presented in \cite[][cosmic shear]{Asgari2021_AA_645_104}, \cite[][$3\times2$pt]{Heymans2021_AA_646_140} and \cite[][beyond $\Lambda$CDM]{Troster2021_AA_649_88}, with the methodology presented in \cite{Joachimi2021_AA_646_129}.

Funding for SDSS-III has been provided by the Alfred P. Sloan Foundation, the Participating Institutions, the National Science Foundation, and the U.S. Department of Energy Office of Science. The SDSS-III web site is \url{http://www.sdss3.org/}.

SDSS-III is managed by the Astrophysical Research Consortium for the Participating Institutions of the SDSS-III Collaboration including the University of Arizona, the Brazilian Participation Group, Brookhaven National Laboratory, Carnegie Mellon University, University of Florida, the French Participation Group, the German Participation Group, Harvard University, the Instituto de Astrofisica de Canarias, the Michigan State/Notre Dame/JINA Participation Group, Johns Hopkins University, Lawrence Berkeley National Laboratory, Max Planck Institute for Astrophysics, Max Planck Institute for Extraterrestrial Physics, New Mexico State University, New York University, Ohio State University, Pennsylvania State University, University of Portsmouth, Princeton University, the Spanish Participation Group, University of Tokyo, University of Utah, Vanderbilt University, University of Virginia, University of Washington, and Yale University.

\section*{Data Availability}

The {\sc Aemulus} and UNIT simulations used in this article are publicly available at \url{https://aemulusproject.github.io/} and \url{http://www.unitsims.org/}, respectively. The DES, KiDS, and SDSS data sets analysed are available at \url{https://www.darkenergysurvey.org/}, \url{http://kids.strw.leidenuniv.nl/}, and \url{https://www.sdss.org/}, respectively. All derived data generated in this research as well as code used will be shared on reasonable request to the corresponding author.

\bibliographystyle{mnras}
\bibliography{bibliography}

\appendix

\section{Gaussian Process modelling}
\label{sec:gp_modelling}

Our analysis method requires generalising the dependence of the maximum likelihood or evidence as a function of cosmology for the $40$ {\sc Aemulus} simulations to arbitrary cosmologies. By default, we utilise the method described in section \ref{subsec:cosmology_fitting} that fits the results for the $40$ simulations with a multi-dimensional skew-normal distribution in the ($S_8$, $\Omega_{\rm m}$, $w$)-plane. Here, we test an alternative approach using Gaussian Process (GP) emulation. This follows similar applications of GP emulation in the literature \citep{Zhai2019_ApJ_874_95, Yuan2020_MNRAS_493_5551}. The main difference to the aforementioned works is that we are emulating only a single summary statistic, $\mathcal{L}_{\rm max}$, as a function of cosmology instead of all observables as a function of galaxy and cosmology parameters \citep{Lange2019_MNRAS_490_1870}.

We apply the alternative GP emulation technique to the mock analyses described in section \ref{sec:mocks}. The cosmological posterior is based on $\mathcal{L}_{\rm max}$, similar to the results in Fig.~\ref{fig:posterior_m_l} and the cosmological parameters considered are $S_8$, $\Omega_{\rm m}$ and $w$. GP interpolation is performed using the publicly available {\sc GPy} package\footnote{\url{https://github.com/SheffieldML/GPy/}}. To train a GP one needs to first determine the kernel that best constrains the covariance matrix of the training set. We first perform a $k$-fold cross validation over a set of kernels such as polynomial, exponential, RBF, Mat\'ern $3/2$ and Mat\'ern $5/2$ to determine the kernel with maximum predictive power. $k$-fold cross validation involves splitting the data set, the $40$ simulations with their respective cosmological parameters and $\mathcal{L}_{\rm max}$ values, into $k$ equal-sized groups. Afterwards, each of the $k = 8$ groups is used as a test set after training the GP on the remaining data. This allows us to empirically asses the predictive power of different kernels. Of all the kernels, we find the Mat\'ern $5/2$ to give the best performance.

\begin{figure}
    \includegraphics{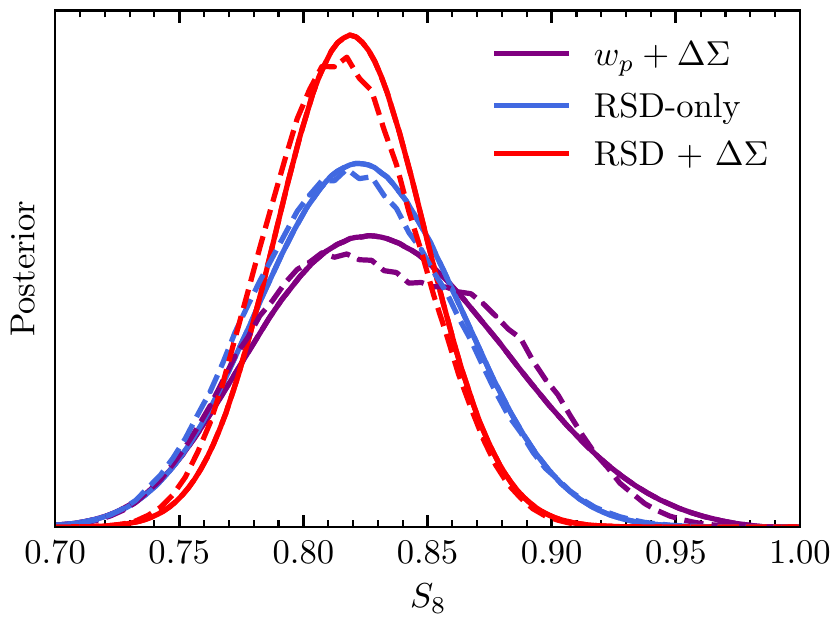}
    \caption{Inferred posterior constraints on $S_8$ from the unmasked mock catalogues. We compare the results from the default procedure to model the likelihood as a function of cosmology (solid) using skew-normals to an alternative approach employing Gaussian Process fitting (dashed).}
    \label{fig:gp}
\end{figure}

After determining the best kernel, we train the GP on the entire set of $40$ simulations and use the interpolated $\mathcal{L}_{\rm max}$ as our proxy for the cosmological posterior. We note that, contrary to the default analysis, this procedure does not account for the impact of uncertainties in the simulation predictions on the final cosmology posterior. However, this effect was found in \cite{Lange2022_MNRAS_509_1779} to be negligible when analysing the mocks with {\sc Aemulus} simulations. In Fig.~\ref{fig:gp}, we compare the inferred posteriors on $S_8$ against the results obtained from the default analysis procedure involving skew-normals. Overall, we find both approaches to give highly consistent results, providing further evidence for the robustness of our analysis method.

\section{Galaxy--halo connection}
\label{sec:galaxy-halo-connection}

\begin{figure*}
    \includegraphics{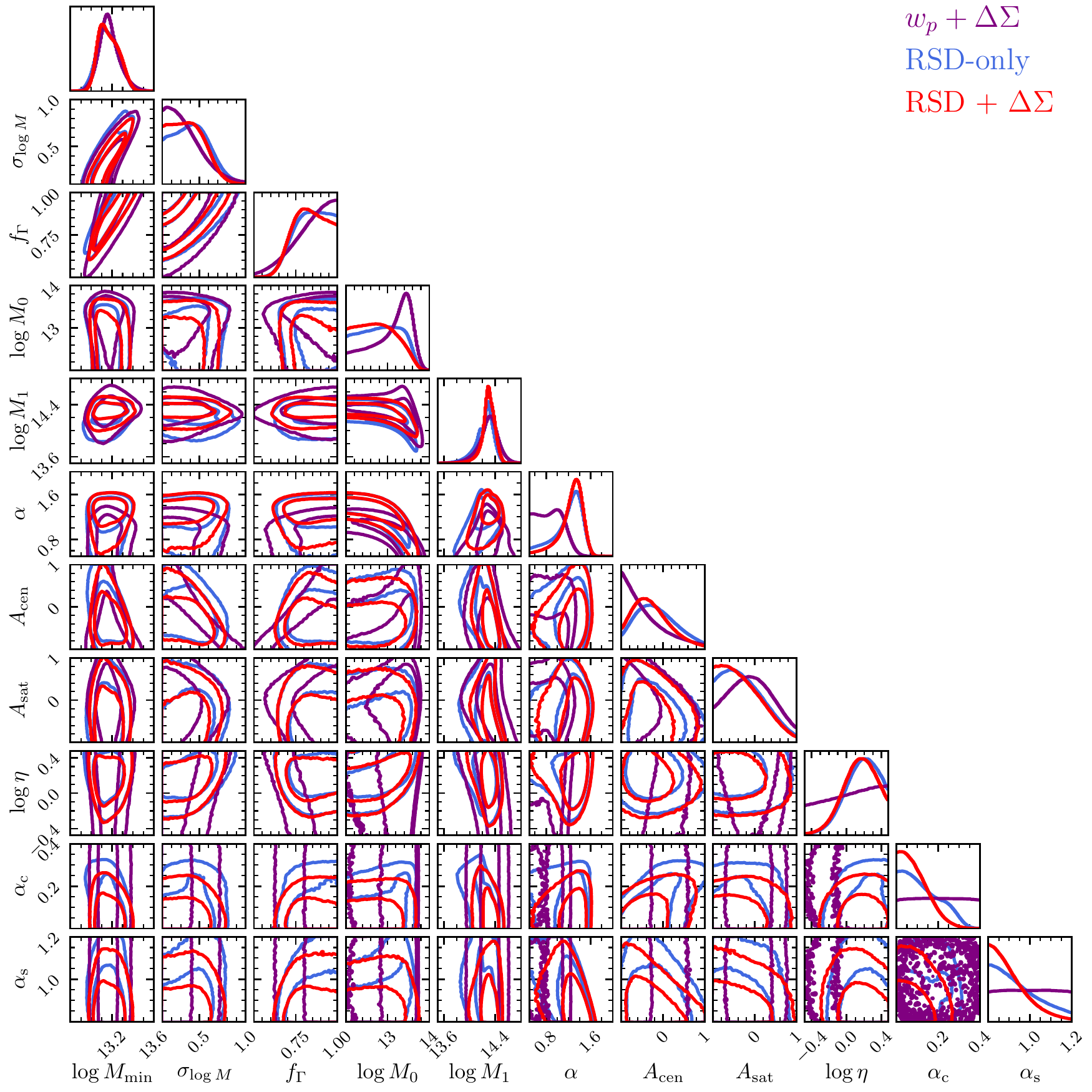}
    \caption{Posterior constraints on galaxy–halo connection parameters for the $0.18 \leq z < 0.30$ sample in the SGC after marginalisation over cosmology. We show constraints coming from redshift-space clustering (blue), the combination of projected clustering and lensing (purple) and RSDs combined with lensing (red). Contours denote $68$ and $95\%$ confidence regions.}
    \label{fig:1s_full_posterior}
\end{figure*}

Here, we present and discuss posterior constraints on the galaxy--halo connection $\mathcal{G}$. Since we fit each of the $40$ simulations to data, we naturally get $40$ posterior constraints $P(\mathcal{G} | \mathcal{C}_i)$ for $40$ different cosmologies $\mathcal{C}_i$. We combine these $40$ by taking the average weighted by the profile likelihood each a simulation,
\begin{equation}
    P(\mathcal{G}) = \frac{\sum P(\mathcal{G} | \mathcal{C}_i) \mathcal{L}_p (\mathcal{C}_i)}{\sum \mathcal{L}_p (\mathcal{C}_i)} \, .
\end{equation}
As an example, we show in Fig.~\ref{fig:1s_full_posterior} all one and two-dimensional posteriors on the galaxy--halo connection parameters for the $0.18 \leq z < 0.30$ sample in the SGC. Similar to the results for cosmology as well as central velocity and assembly bias discussed below, this figure indicates good agreement between the constraints derived from redshift-space clustering, the combination of projected clustering and lensing and RSDs combined with lensing. This figure also demonstrates that the addition of gravitational lensing as a constraint does not add significant constraining power on galaxy--halo connection parameters compared to redshift-space clustering alone. This is expected since we only include gravitational lensing in the two-halo regime. At fixed cosmology, gravitational lensing in this regime primarily contains information on the large-scale bias, something already contained within clustering measurements.

Our posterior constraints on the HOD parameters, particularly $M_{\rm min}$, $M_1$ and $\sigma_{\log M}$, present significant variation amongst the different galaxy samples studied in this work. This is expected since the different samples represent galaxy populations at different redshifts, stellar masses etc. In this appendix we focus on velocity bias and assembly bias, as we find that the conclusions drawn below apply equally well to each of the galaxy samples we consider.

\begin{figure*}
    \includegraphics{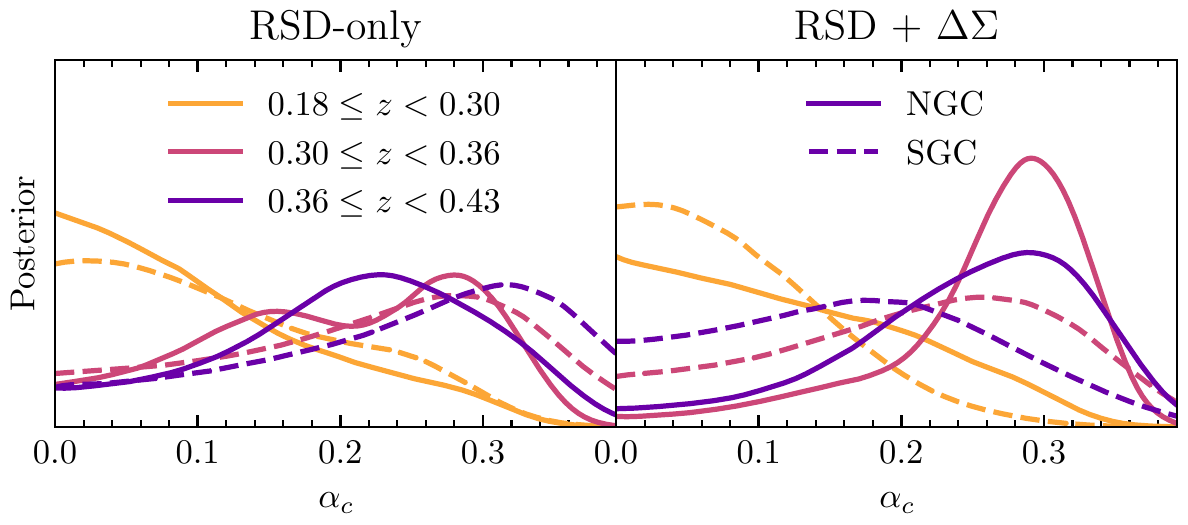}
    \caption{Posterior constraints on central velocity bias after marginalisation over cosmology. Different panels indicate the observational constraints used: redshift-space clustering (left) and the combination of redshift-space clustering and galaxy--galaxy lensing (right). Different lines indicate different samples.}
    \label{fig:alpha_c}
\end{figure*}

In Fig.~\ref{fig:alpha_c}, we present the posterior constraints on $\alpha_c$, the central velocity bias parameter. As expected, redshift-space clustering is able to put some constraints on $\alpha_c$, whereas we do not show the combination of projected galaxy clustering and galaxy--galaxy lensing since it is insensitive to this parameter. Overall, our results are consistent with little to no central velocity bias, $\alpha_c = 0$, and agree with the findings in \cite{Lange2022_MNRAS_509_1779}. Similarly, our results here do not contradict the findings in \cite{Guo2015_MNRAS_446_578} where $\alpha_c > 0$. In the present work, we define central velocity with respect to the inner $10\%$ of the halo particles \citep{Behroozi2013_ApJ_762_109} instead of the inner $25\%$ as in \cite{Guo2015_MNRAS_446_578}. The latter definition is expected to imply larger values for $\alpha_c$ \citep{Ye2017_ApJ_841_45}.

\begin{figure*}
    \includegraphics{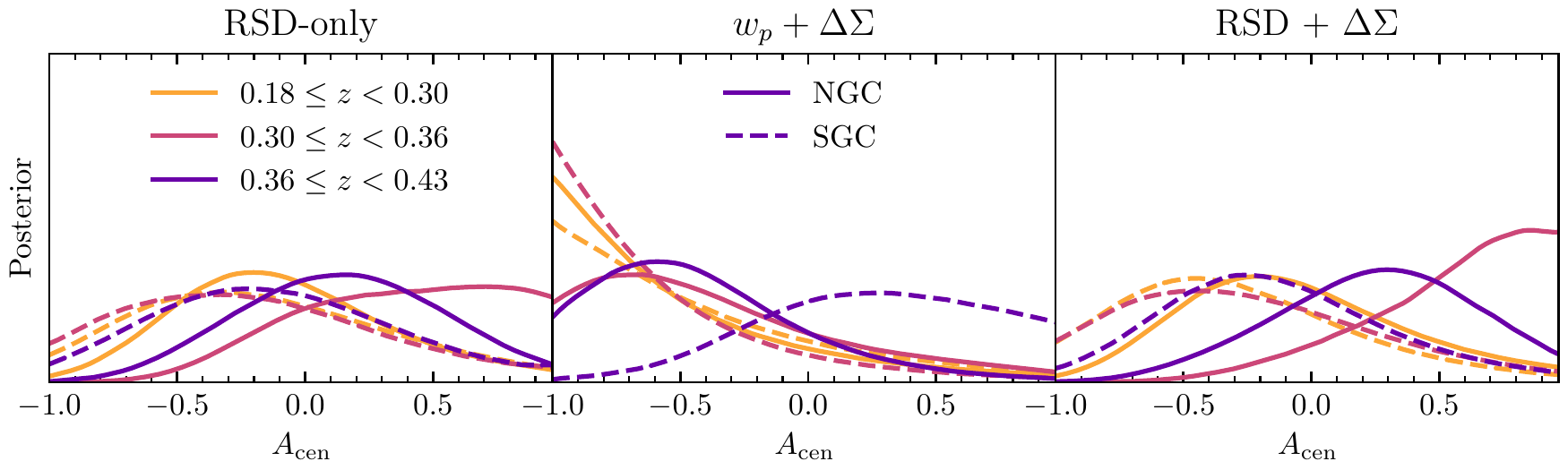}
    \caption{Same as Fig.~\ref{fig:alpha_c} except for focussing on the central assembly bias parameter $A_{\rm cen}$ and also showing results for the combination of projected clustering and galaxy--galaxy lensing (middle).}
    \label{fig:a_cen}
\end{figure*}

Fig.~\ref{fig:a_cen} shows our constraints on the central assembly bias parameter $A_{\rm cen}$. We find no strong constraints on assembly bias and our results are consistent with no assembly bias, $A_{\rm cen} = 0$. As discussed in \cite{Lange2019_MNRAS_490_1870, Lange2022_MNRAS_509_1779}, this is in part due to the degeneracy between $A_{\rm cen}$ and cosmology. At the same time, even at fixed cosmology, we find neither redshift-space clustering nor the combination of projected clustering and galaxy--galaxy lensing to be very sensitive to assembly bias. Similarly, our analysis also does not yield any noteworthy constraints on the satellite assembly bias parameter $A_{\rm sat}$, either. Other summary statistics beyond two-point correlation functions might be needed to robustly constrain assembly bias \citep[see e.g.][]{Wang2019_MNRAS_488_3541, StoreyFisher2022_arXiv_2210_3203}.

\label{lastpage}

\end{document}